\documentstyle[12pt]{article}
\begin{document}
\title{{\bf LOW VELOCITY
\\ GRAVITATIONAL CAPTURE
\\ BY LONG COSMIC STRINGS}
\thanks{Alberta-Thy-01-99, gr-qc/9902038}}
\author{
Don N. Page
\thanks{Internet address:
don@phys.ualberta.ca}
\\
CIAR Cosmology Program, Institute for Theoretical Physics\\
Department of Physics, University of Alberta\\
Edmonton, Alberta, Canada T6G 2J1
}
\date{(1999 February 11)}

\maketitle
\large

\begin{abstract}
\baselineskip 14.8 pt

	Coupled ordinary differential equations are derived
for the distant gravitational interaction
of a compact object of mass $M$ and charge $Q$
with an initially straight, infinitely long, cosmic string
of tension $\mu \ll 1/G \equiv 1$,
when the relative velocities are
very low compared to the speed of light $c \equiv 1$.
(An intermediate result of this derivation
is that any localized force ${\mathbf F}(t)$
on the string that is confined
to a single plane perpendicular
to the initial string configuration
gives the intersection of the string with this
plane---the point where the force is applied---the
velocity ${\mathbf F}(t)/(2\mu)$.) 
The coupled equations are then used to calculate
the critical impact parameter $b_{\rm crit}(v_0)$
for marginal gravitational capture
as a function of the incident velocity $v_0$.
For $v_0 \ll (1-Q^2/M^2)^{1/3}\mu^{2/3}$,
so that the string acts relatively stiffly,\\
$b_{\rm crit} \approx
{\pi\over 4}\left[12\mu^3
\left(1-{Q^2\over M^2}\right)^4 \right]^{1\over 5}
M v_0^{-{7\over 5}}
+{\pi\over 5}\left[{27\over 16 \mu}
\left(1-{Q^2\over M^2}\right)^2\right]^{1\over 5}
M v_0^{-{1\over 5}} + O\left({M v_0\over\mu}\right)$.\\
For $(1-Q^2/M^2)^{1/3}\mu^{2/3} \ll v_0 \ll 1 - Q^2/M^2$,
so that the string acts essentially
as a test string that stays nearly straight,\\
$b_{\rm crit} \approx
\left[{\pi\over 2}
\left(1-{Q^2\over M^2}\right)\right]^{1\over 2}
M v_0^{-{1\over 2}}
+{\pi\over 4}\left(1-{Q^2\over M^2}\right)
\mu M v_0^{-2} + O(M\mu^2 v_0^{-{7\over 2}})
+ O(M)$.\\
Between these two limits the critical
impact parameter is found numerically
to fit a simple algebraic combination
of these two formulas to better than 99.5\% accuracy.

\end{abstract}
\normalsize
\baselineskip 15 pt
\newpage

\section{Introduction}

\hspace{.20in}	The gravitational interactions
between compact objects and long cosmic strings
have been calculated in several recent papers
\cite{JPD,JPDnew,Page,JPD3}
in the test-string approximation,
in which one assumes that the dimensionless
string tension $\mu$ (using units in which
Newton's gravitational constant is $G=1$)
is much smaller than any other parameter
whose ratio with $\mu$ is relevant.
Then the string obeys the Nambu-Goto equations of motion
\cite{Nambu,Goto,Carter}
in the gravitational field of the object
and has negligible back reaction on
the motion of the object.

	It is indeed a good approximation
for cosmic strings that $\mu \ll 1$,
but I shall show that at low initial relative velocities $v_0$
and near the critical impact parameter,
the ratio $\mu/v_0^{3/2}$ is important,
so the test-string approximation is accurate
only if the velocity is large in comparison
to $\mu^{2/3}$.
For velocities very low compared with this,
the string acts relatively stiffly,
so that the motion of the object responds
more than the motion of the string.

	Here I shall derive
coupled ordinary differential equations
of motion for an initially straight, infinitely long,
cosmic string interacting gravitationally with
a compact object of mass $M$ and charge $Q$
(i.e., one whose size is negligible
in comparison with the distance
between the object and the string),
using the approximations $\mu \ll 1$
and $v_0 \ll 1$ but allowing $\mu/v_0^{3/2}$
to be arbitrary.
The low velocity allows the part of the string
nearest the object to remain nearly straight,
and then the force ${\mathbf F}$
on the string is a simple function
of the string-object separation and relative velocity.
On a much larger scale this force is nearly
localized and hence gives the point on the string
nearest the object the velocity ${\mathbf F}/(2\mu)$,
as we shall see.

	For fixed $M$, $Q$, and $\mu$,
after taking out Poincare transformations
there is a two-parameter set of solutions
that can be characterized by the initial
speed $v_0$ of the object relative to the string
(when they are infinitely far away in the infinite past)
and by the impact parameter $b$
(how far the object would miss the string
if both moved uniformly with no interactions).
For each $v_0$ there is a critical impact parameter,
$b_{\rm crit}$, such that if $b < b_{\rm crit}$
the object becomes gravitationally captured
and bound to the string,
but if $b > b_{\rm crit}$,
the object and string scatter
and asymptotically move freely of each other.
In this paper the coupled equations of motion
of the object and the string will be solved
to give the function $b_{\rm crit}(v_0)$,
analytically when the ratio $\mu/v_0^{3/2}$
is either very large or very small,
and numerically otherwise.
In the latter case simple algebraic combinations
of the analytic results at the two extremes
reproduce the numerical results over the entire range
to within 1\%, or even within 0.43\%
if two arbitrary exponents are chosen appropriately.

\section{Gravitational force between an object
and a straight string}

\hspace{.20in}	Strictly speaking,
the gravitational interaction between
an object and a string should be done
in terms of curved spacetime.
However, when $\mu \ll 1$
and when one is in the weak-field regime of the object
(which is where the string is, under the assumption
that the object size is much smaller
than its distance to the string),
one can do an analysis in terms of fictitious forces
acting in a fictitious flat background spacetime,
just as one does for Newtonian gravity
(though one need not assume that
the gravitational field of the string
is Newtonian or that the object velocity is low to do this).

	We assume that the string is infinitely long and is
nearly straight in the region where it is fairly near
its minimum separation from the object,
with no oscillations coming in along the string from infinity.
Given the assumed initial straight, static configuration
of the string, this assumption of approximate straightness
of the part of the string nearest the object for all times
is justified by the assumption of low transverse velocities,
since the fact that the disturbances of the string
propagate up and down it, away from their source,
at the speed of light, means that the length of string
that is disturbed is much greater than the transverse
disturbance, so that the angular bending of the string
is small, even if the linear bending becomes large.

	Then we can go to the frame in which the part
of the string nearest the object is at rest along
the $z$-axis and the object is in the $x$-$y$ plane
with velocity of magnitude $v_r$
in the positive $x$-direction,
the relative velocity of the object with respect
to the piece of string nearest the object.
To linear order in $M$ and $Q$, the gravitational force
of the string on the object is zero when $v_r=0$.
However, the string produces a conical spacetime
with a deficit angle $2\pi(4\mu)$ \cite{Vil},
so when this is flattened out in the fictitious flat
background spacetime with azimuthal angle $\varphi$
that ranges from 0 to $2\pi$,
a moving object appears to be bending toward
the string at the rate of $4 \mu$
times the rate at which $\varphi$ is changing,
i.e., at the rate of $4\mu\dot{\varphi} = 4\mu v_r y/r^2$,
where
 \begin{equation}
 r = \sqrt{x^2 + y^2} 
 \label{eq:1}
 \end{equation}
is the distance from the object to the string
(ignoring relative corrections of order $\mu$
that depend on how the conical spacetime
is flattened out to produce the fictitious flat spacetime).

	When this bending rate is multiplied by the
momentum $p = M \gamma_r v_r$ of the object,
where $\gamma_r = 1/\sqrt{1-v_r^2}$, 
one gets the part of the transverse fictitious force
on the object that is linear in $M$,
$4\mu M \gamma_r v_r^2 y/r^2$.
The longitudinal fictitious force
(in the direction of the object velocity ${\mathbf v}_r$
in the instantaneous frame of the piece of string
closest to the object,
using boldface symbols to denote spatial vectors
in the two-dimensional plane perpendicular
to the string in the fictitious flat spacetime)
that is linear in $M$ depends on the flattening procedure
used to generate the fictitious flat metric.
One can choose it so that the total fictitious
force on the object that is linear in $M$ is
 \begin{equation}
 {\mathbf F}_{\rm{linear}}
 = -4\mu M\gamma_r v_r^2 {\mathbf r}/r^2.
 \label{eq:2}
 \end{equation}
By Newton's third law or conservation of momentum,
there will be an equal and opposite
fictitious force on the string
to describe its motion in the fictitious
flat spacetime.

	To explain and justify this fictitious force
in greater detail,
consider the geodesic motion of a test mass $M$
in the conical spacetime of a cosmic string \cite{Vil},
 \begin{eqnarray}
 ds^2 &=& - dt^2 + dz^2 + d\rho^2 + (1-4\mu)^2 \rho^2 d\varphi^2
	\nonumber \\
	&=& - dt^2 + dz^2 + r^{-8\mu}(dr^2 + r^2 d\varphi^2)
	= - dt^2 + dz^2 + e^{2\phi}(dr^2 + r^2 d\varphi^2),
 \label{eq:3}
 \end{eqnarray}
where
 \begin{equation}
 r = [(1-4\mu)\rho]^{1/(1-4\mu)}
 \label{eq:4}
 \end{equation}
and
 \begin{equation}
 \phi = -4\mu \ln{r}
 \label{eq:5}
 \end{equation}
(not to be confused with the azimuthal angle $\varphi$).

	One can go to the frame in which the object
moves in the plane $z=0$, in which case the motion
is in an ultrastatic $(2+1)$-dimensional spacetime
with conformally flat spatial sections,
whose general metric form is
 \begin{equation}
 ds^2 = - dt^2 + e^{2\phi}\delta_{ij} dx^i dx^j.
 \label{eq:6}
 \end{equation}
Timelike geodesics of such metrics have
constant
 \begin{equation}
 E = M{dt \over d\tau}
 \label{eq:7}
 \end{equation}
and constant
 \begin{equation}
 P^2 = M^2 e^{2\phi}\delta_{ij}{dx^i \over d\tau}{dx^j \over d\tau}
     = E^2 - M^2.
 \label{eq:8}
 \end{equation}

	One can also readily show
that the spacetime geodesic equation
implies that the spatial trajectory is itself a geodesic
of the spatial metric $e^{2\phi}\delta_{ij} dx^i dx^j$
and obeys the equation
 \begin{equation}
 {d^2 x^i \over d\sigma^2}
 = \left(\delta^{ij} - {dx^i \over d\sigma}
	{dx^j \over d\sigma}\right)
	\phi_{,j},
 \label{eq:9}
 \end{equation}
where $\sigma$ is the spatial distance along the trajectory,
 \begin{equation}
 d\sigma^2 = \delta_{ij} dx^i dx^j,
 \label{eq:10}
 \end{equation}
in the fictitious flat spacetime metric
 \begin{equation}
 ds_{\flat}^2 = - dt_{\flat}^2 + \delta_{ij} dx^i dx^j.
 \label{eq:11}
 \end{equation}

	One can reproduce the spatial geodesic
of the conformally-flat spatial metric
$e^{2\phi}\delta_{ij} dx^i dx^j$
by a fictitious spatial force
 \begin{equation}
 {\mathbf F} = {d{\mathbf p}\over dt_{\flat}}
 = M \gamma_r v_r^2 {\mathbf \nabla}\phi + f {\mathbf v}_r
 \label{eq:12}
 \end{equation}
in the fictitious flat spacetime, where
 \begin{equation}
 {\mathbf p} = M \gamma_r {\mathbf v}_r
 = M \gamma_r d{\mathbf r}/dt_{\flat}
 \label{eq:13}
 \end{equation}
is the fictitious flat spacetime momentum
of the particle of mass $M$,
of flat spacetime position vector ${\mathbf r}$
(with components $x^i$ that are raised and lowered
by the fictitious flat spatial metric
$\delta_{ij} dx^i dx^j$ in this fictitious force analysis),
of spatial velocity ${\mathbf v}_r = d{\mathbf r}/dt_{\flat}$
with magnitude
$v_r = \sqrt{{\mathbf v}_r\cdot{\mathbf v}_r}
 = \sqrt{\delta_{ij} v_r^i v_r^j}$,
and of $\gamma_r = 1/\sqrt{1-v_r^2}$.

	In the expression above for the fictitious force
${\mathbf F}$, one can have an arbitrary function $f$
multiplying the velocity ${\mathbf v}_r$ if one just wants
to reproduce the spatial trajectory of the geodesic
of the actual metric (\ref{eq:6}).
If one also wants to reproduce the time dependence
of the trajectory, so $x^i(t_{\flat})$
under the fictitious force in the fictitious metric
has the same dependence on $t_{\flat}$
as $x^i(t)$ does on $t$ for the geodesic
in the actual metric, then one needs
 \begin{equation}
 f = -M \gamma_r(\gamma_r^2 + 1)
	{\mathbf v}_r\cdot{\mathbf \nabla}\phi.
 \label{eq:14a}
 \end{equation}

	However, for the purposes of this paper,
the precise time dependence of the trajectory
of an object passing a cosmic string
is not so important as the trajectory itself,
so I shall set $f = 0$.
Then if one uses the form of $\phi$
given by Eq. (\ref{eq:5}) for the cosmic string,
one gets the fictitious force given by Eq. (\ref{eq:2}).

	To quadratic order in $M$ and $Q$,
A. G. Smith \cite{Smi} calculated from the
effect of the deficit angle on the
three-dimensional Laplacian for
the Newtonian gravitational and electrostatic
fields of the object that the string exerts
an attractive force (if $M^2 > Q^2$)
of magnitude $\pi\mu(M^2 - Q^2)/(4r^2)$
on the object when the relative velocity $v_r$
can be neglected.
(When $v_r$ is not negligible in comparison with unity,
the quadratic force is overwhelmed by the linear
force, so one does not need the high-velocity
corrections to the quadratic force.)
In two-dimensional vectorial form,
 \begin{equation}
 {\mathbf F}_{\rm{quadratic}}
 = -\pi\mu (M^2 - Q^2){\mathbf r}/(4r^3).
 \label{eq:14}
 \end{equation}

	The conservation of momentum
implies that the object exerts an equal
but opposite force on the string,
and this indeed gives precisely the bending
of a distant static string
in the Reissner-Nordstrom metric \cite{who1}.

	When we add the fictitious forces
that are linear and quadratic in $M$ and $Q$,
we get a total fictitious force on the object that,
in the limit of a small dimensionless
string tension $\mu \ll 1$
and in the limit of a large separation
$r \gg M$, is approximately
 \begin{equation}
 {\mathbf F}_M
 = - {4\mu M\gamma_r v_r^2 {\mathbf r}\over r^2}
    - {\pi\mu (M^2 - Q^2){\mathbf r}\over 4r^3}.
 \label{eq:15}
 \end{equation}
The first term is valid for all velocities ${\mathbf v}_r$
but neglects additional terms
in the direction of ${\mathbf v}_r$,
and the second term, which is only important
if the velocity is very small, neglects
corrections when $v_r$ is not small.

\section{Motion of a long string
with a localized force}

\hspace{.20in}	In the fictitious flat spacetime,
the response of the compact object
to the fictitious force is simply the
relativistic version of Newton's second law,
the first equality of Eq. (\ref{eq:12}).
However, the response of the string
to the equal and opposite fictitious
force needs to be worked out.

	As noted above, the low-transverse-velocity
assumption means that the length of string
that is bent (which develops at the speed of light)
is much greater than the transverse bending
(which develops at the transverse velocity
of the string, much less than the speed of light
by assumption).  Hence, one can look at the bending
on a scale much larger than that of the transverse
separation between the string and the object
and yet much smaller than that of the length
of string that is bent.  On this scale, the force
on the string appears to be localized
at the position on the string nearest the object
and is given to good approximation by Eq. (\ref{eq:15})
in terms of the string-object transverse separation
${\mathbf r}$ and the relative velocity $v_r$.

	Let us do a general analysis of the effect
of such a localized force on an initially static, straight
cosmic string, without, in the general analysis,
assuming that the transverse velocity
of the string is small, even though in our
particular application we need this assumption
to justify our assumption that the force
is effectively local and is given by Eq. (\ref{eq:15}).

	Suppose a time-dependent horizontal force
in the $x$-$y$ plane is applied at $z=0$
to an infinite string initially lying at rest
along the $z$-axis (vertically), with no other forces on it.
This force causes equal horizontal disturbances
to go up (for $z>0$) and down (for $z<0$)
the string at the speed of light
(propagating away from where they are generated
by the time-dependent force, at $z=0$;
assuming that the infinite string was initially
at rest and has no forces on it at $z \neq 0$
implies that no disturbances are coming in to $z=0$):
 \begin{equation}
 {\mathbf r}_s(t,z)
 = \theta(z){\mathbf r}_{\bot}(t-z)
 + \theta(-z){\mathbf r}_{\bot}(t+z),
 \label{eq:16}
 \end{equation}
where $\theta(z)$ is the Heaviside step function of $z$
and each of the two occurrences of
${\mathbf r}_{\bot}$ represents the same
function of the argument inside the following
bracket (either $t-z$ for the first term,
which is nonzero for $z>0$ where $\theta(z)=1$
and $\theta(-z)=0$,
or $t+z$ for the second term,
which is nonzero for $z<0$ where $\theta(-z)=1$
and $\theta(z)=0$).

	Focus on the infinitesimal piece of string
just above $z=0$ at some arbitrary time $t$.
Its tilt or slope away from vertical is
 \begin{equation}
 {\partial {\mathbf r}_s(t,z) \over \partial z}
 = - {\partial {\mathbf r}_s(t,z) \over \partial t}
 = - \dot{{\mathbf r}}_{\bot},
 \label{eq:17}
 \end{equation}
where the overdot in the last term denotes
a derivative with respect to the unlisted argument
$t-z$ evaluated at $z=0$.
Denote the magnitude of this slope by
 \begin{equation}
 v_s \equiv |\dot{{\mathbf r}}_{\bot}|
    \equiv \sqrt{\dot{{\mathbf r}}_{\bot}\cdot
	\dot{{\mathbf r}}_{\bot}}
 = \tan{\psi},
 \label{eq:18}
 \end{equation}
where $\psi$ is the instantaneous angle of the tilt
from the vertical.

	This piece of string can be considered to
be moving straight upward at the speed of light (unity),
the velocity at which the pattern moves for $z>0$,
but locally any component of the velocity
parallel to the string has no effect.
To calculate the energy and momentum
carried by a piece of string,
what is relevant is the transverse velocity,
the component perpendicular to the string.
Here the infinitesimal piece of string
is tilted at an angle $\psi$ from the vertical,
so the transverse velocity is tilted
at an angle $\pi/2 - \psi$ from the
vertical direction along which the pattern speed is unity.
Therefore, the magnitude of the transverse velocity is
 \begin{equation}
 v_t = \sin{\psi} = {v_s \over \sqrt{1+v_s^2}},
 \label{eq:19}
 \end{equation}
and the associated relativistic $\gamma$
factor is
 \begin{equation}
 \gamma_t = {1 \over \sqrt{1-v_t^2}}
 = \sec{\psi} = \sqrt{1+v_s^2}.
 \label{eq:20}
 \end{equation}

	The length of this infinitesimal piece of string is
 \begin{equation}
 dl = (\sec{\psi}) dz = \sqrt{1+v_s^2}\: dz,
 \label{eq:21}
 \end{equation}
and its energy is
 \begin{equation}
 dE = \mu \gamma_t dl = \mu (\sec^2{\psi}) dz
 = \mu (1+v_s^2) dz.
 \label{eq:22}
 \end{equation}
The transverse momentum
(at the angle $\psi$ from the horizontal $x$-$y$ plane)
is the transverse velocity times the energy,
with magnitude
 \begin{equation}
 dp_t = v_t dE = \mu (\sin{\psi}\sec^2{\psi}) dz
 = \mu v_s \sqrt{1+v_s^2}\: dz.
 \label{eq:23}
 \end{equation}
The vertical component of this momentum is
 \begin{equation}
 dp_z = \sin{\psi}dp_t = \mu (\tan^2{\psi}) dz
 = \mu v_s^2 dz,
 \label{eq:23b}
 \end{equation}
and the horizontal component has magnitude
 \begin{equation}
 dp_h = \cos{\psi}dp_t = \mu (\tan{\psi}) dz
 = \mu v_s dz
 \label{eq:24}
 \end{equation}
and is in the direction of $\dot{{\mathbf r}}_{\bot}$.

	This infinitesimal piece of the displaced-string
pattern that is moving upward at the speed of light
from the force at $z=0$
takes a time $dt = dz$ to be generated.
A mirror-image pattern is moving
downward at the speed of light
from the force at $z=0$,
and during the same infinitesimal time
an infinitesimal piece of string
with the same magnitude
of $dz$ is generated moving downward
just below $z=0$.
The vertical components of these two
pieces are opposite and hence cancel,
but the horizontal components are in
the same direction and so add to make
the two infinitesimal pieces of string
generated during the time $dt$
have a total momentum of
 \begin{equation}
 d{\mathbf p} = 2 \mu \dot{{\mathbf r}}_{\bot} dt
 = 2 \mu d{\mathbf r}_{\bot}
 = 2 \mu d{\mathbf r}_s,
 \label{eq:24b}
 \end{equation}
where here and henceforth I shall use
${\mathbf r}_s$ for ${\mathbf r}_s(t,0)$
at $z=0$, the horizontal position
of the string in the $x$-$y$ plane
where the force is applied.

	Since the string was assumed to start
off from rest along the $z$-axis,
at ${\mathbf r}_s(t,z) = 0$ for large negative $t$,
the total momentum of the string is simply
 \begin{equation}
 {\mathbf p}_s = 2 \mu {\mathbf r}_s,
 \label{eq:25}
 \end{equation}
and the force on the string is
 \begin{equation}
 {\mathbf F}_s = {d{\mathbf p}_s \over dt}
 = 2 \mu \dot{{\mathbf r}}_s.
 \label{eq:26}
 \end{equation}

	The string will react back
with an equal and opposite force on whatever
is forcing it to move horizontally.
It is interesting that this reaction force
is precisely an idealized friction force
directly proportional to velocity.
The fact that momentum is carried away
along the string at the speed of light,
never to return
(since the string is assumed to be infinite),
is apparently what allows an initially straight,
infinitely long string to act as a perfect source of friction.

	One application of Eq. (\ref{eq:25})
is to give a very simple derivation of the
net displacement of an initially straight
string by a gravitating object of mass $M$
that passes by with large impact parameter
$b \gg M/v_0^2$, so that the fictitious force term linear
in $M$, the first term of Eq. (\ref{eq:15}),
dominates over the force term quadratic in $M$ and $Q$,
the second term of Eq. (\ref{eq:15}).
Then the object trajectory bends by half the deficit angle,
i.e., by an angle $4\pi\mu \ll 1$, so its momentum
changes by
 \begin{equation}
 \Delta p = 4\pi\mu p = 4\pi\mu M \gamma_0 v_0.
 \label{eq:27}
 \end{equation}
By Newton's third law, the string's momentum changes
by the opposite amount, so Eq. (\ref{eq:25})
implies that the string gets displaced by
 \begin{equation}
 \Delta r = \Delta p/(2\mu) = 2\pi M \gamma_0 v_0,
 \label{eq:28}
 \end{equation}
just as De Villiers and Frolov found \cite{JPDnew}.

	If one inverts Eq. (\ref{eq:26}) to get
 \begin{equation}
 \dot{{\mathbf r}}_s = {{\mathbf F}_s \over 2 \mu},
 \label{eq:29}
 \end{equation}
one notices the curious fact that apparently
a force larger than twice the string tension $\mu$
can move a string faster than the speed of light,
seemingly a violation of special relativity,
despite the fact that a completely
(special) relativistic analysis was used
in the derivation of the relationship
between horizontal force and velocity.
However, one can make two comments
about this paradox:

	First, for all finite horizontal speeds $v_s$
of the string, even `superluminal' $v_s>1$,
the physical transverse speeds $v_t$ of the string,
given by Eq. (\ref{eq:19}) above, are less than
the speed of light.  The horizontal speed $v_s$
of the kink in the string at $z=0$ where the force
is applied is somewhat analogous to the speed
of the point where the two blades of a pair
of scissors intersect.  Even though the two
blades are constrained to move slower than
the speed of light, there is nothing
in special relativity that constrains
the intersection point to move slower than light.

	One might object that the scissors
are more nearly analogous to a pair of
intersecting strings both moving in the same plane,
whose intersection point can indeed move 
faster than light (at least if one assumes
no interactions at the intersection,
so that the two strings each move freely).
In this case it seems to be the prior existence
of each of the two strings on both sides
of the intersection that allows
the intersection to move faster than light
while neither string does so.
However, for the single string being considered
in this paper, whose kink at $z=0$
moves horizontally faster than light
if the force is greater than twice the string tension,
there is no pre-existing string beyond the kink,
so it seems that something physical must be moving
faster than light.

	The partial answer to this worry
is that for a horizontal force greater than $2\mu$,
string is being created faster than light,
but it is not being moved faster than light
(in the transverse direction, the only motion
that has a physical meaning for
a longitudinal-boost-invariant string
of the type being considered here).
This is somewhat analogous to the possible
nearly-instantaneous freezing of a lake,
in which the water-ice boundary can
in principle move faster than the speed of light
as ice is created (but not moved) faster than light.

	Another analogy would be electron-positron
pair creation in a hypothetical imploding cylindrical
electromagnetic wave that builds up
a huge longitudinal electric field,
say between two capacitor plates
at the ends of the cylinder.
The pair creation can in principle discharge
the plates much faster than the time
taken for light to go from one plate to the other.
If one draws the Feynman diagrams of the
pair creation and annihilation that accomplish
this process, with positrons pictured as electrons
moving backward in time, one can get electron
lines that zigzag forward and backward between
pair creations and annihilations and have
a net spatial motion faster than light between
the two plates. 
One might call this a tachyonic lightning bolt.
And yet there would be no violation
of special relativity with signals traveling faster than light
in this hypothetical process,
since it would be the imploding electromagnetic wave
that locally causes the process, with no causes
traveling faster than the speed of light.

	Considering the creation of string
faster than the speed of light leads on to the second
comment, which is that although 
a string with a kink having a horizontal velocity $v_s>1$
does not seem to violate
the principles of causality in special relativity
(signals not traveling faster than light),
it does seem to be associated with
an instability analogous to tachyonic behavior.
The example of the nearly-instantaneous freezing
of a lake illustrates the instability of the liquid phase
of water at low temperatures, and the example of
the pair creation in the imploding electromagnetic field
illustrates the instability of the zero-current
`vacuum' of the Dirac electron-positron field
in the presence of a strong electric field.
Similarly, tachyonic negative
mass-squared terms in relativistic wave equations,
such as the Klein-Gordon equation,
do not lead to violations of causality
(since disturbances do not propagate faster than light
whatever the sign of the mass-squared term)
but instead lead to instabilities of exponential growth.

	In the case of applying a strong force to a string,
this can lead to an instability in the production of strings.
For example, suppose the force is that of a gauge field
on a massless charged particle attached to the string
at the kink at $z=0$.
(If the particle were massive, its inertia would prevent
the kink from moving faster than light,
so I shall assume the contrary to continue the argument.)
But then if the force were greater than $\mu$,
the energy that a pair of oppositely charged
massless particles extract from the gauge field in being created and
separated would be greater than the energy
needed to create the string joining them.
This would lead to the rapid pair production
of pieces of string with these oppositely charged
gauge particles at the ends in the strong gauge field,
an instability that would presumably rapidly discharge
the gauge field until the force on a charged particle
would be no greater than $\mu$.

	Therefore, if the force is that of a stable gauge field
applied to a massless charged particle on the string,
it appears that the maximum force is the string tension
$\mu$, and therefore the maximum horizontal velocity
of the string kink where the particle is attached is one-half
the speed of light
(relative to the distant parts of the string that are still
at rest along the $z$-axis according to the assumption
that the entire string was originally at rest along
the $z$-axis and that the force is applied only at $z=0$).

\newpage

\section{Coupled motion of a slowly moving compact
object and a nearly straight infinitely long string}

\hspace{.20in}	Now let us combine
the expression (\ref{eq:15})
for the force between a compact object and a distant
nearly straight long string with Newton's second
law for the response of the object
and with Eq. (\ref{eq:29}) for the response of the string,
when the relative velocities are all small compared
with the speed of light, so that Eq. (\ref{eq:15})
indeed applies for giving the force
and so that the force is effectively localized
(relative to the length of string that is bent)
in order for Eq. (\ref{eq:29}) to apply as well.
Use a fictitious flat spacetime
metric with Cartesian coordinates chosen so that
the string is initially at rest along the $z$-axis
and the compact object moves in the $x$-$y$ plane
($z=0$),
and use boldface letters to denote vectors
in this plane.
(The same letters when not bold will denote
the magnitudes of the corresponding boldface vectors.)
In particular, let ${\mathbf r}_M$
denote the position of the compact object
(with mass $M$ and charge $Q$) relative to
the origin $(x,y)=(0,0)$,
let ${\mathbf r}_s$ denote the position of the
string at $z=0$ relative to the origin, and let
 \begin{equation}
 {\mathbf r} = {\mathbf r}_M - {\mathbf r}_s
 \label{eq:30}
 \end{equation}
denote the position of the compact object
relative to the string.

	Then the nonrelativistic
coupled equations of motion
for the compact object and the string become
 \begin{equation}
 M \ddot{{\mathbf r}}_M = - 2 \mu \dot{{\mathbf r}}_s
 = - {\pi\mu(M^2-Q^2){\mathbf r} \over 4 r^3}
 - {4\mu M
	(\dot{{\mathbf r}}\cdot\dot{{\mathbf r}})
	{\mathbf r}\over r^2}.
 \label{eq:31}
 \end{equation}

	It is now convenient to use the unit of time
 \begin{equation}
 T_u \equiv {M \over 2\mu}
 \label{eq:32}
 \end{equation}
and the unit of length
 \begin{equation}
 L_u \equiv
 \left[{\pi\over 16\mu}
	\left(1-{Q^2\over M^2}\right)\right]^{1/3} M,
 \label{eq:33}
 \end{equation}
which we henceforth define to be unity
(unless otherwise specified).
In terms of these units the speed of light is
no longer unity but is
 \begin{equation}
 c
 = \left[{\pi\over 2}\mu^2
	\left(1-{Q^2\over M^2}\right)\right]^{-1/3}
	{L_u \over T_u}
 = \left[{\pi\over 2}\mu^2
	\left(1-{Q^2\over M^2}\right)\right]^{-1/3}
 \gg 1.
 \label{eq:34}
 \end{equation}

	Setting $T_u = L_u = 1$, Eqs. (\ref{eq:31}) become
 \begin{equation}
 \ddot{{\mathbf r}}_M = - \dot{{\mathbf r}}_s
 = - {{\mathbf r} \over r^3}
 - {4\mu(\dot{{\mathbf r}}\cdot\dot{{\mathbf r}}){\mathbf r}
	\over r^2}.
 \label{eq:35}
 \end{equation}

	At an arbitrary time $t_i$,
the freely specifiable initial conditions
are ${\mathbf r}_s(t_i)$, ${\mathbf r}_M(t_i)$,
and $\dot{{\mathbf r}}_M(t_i)$ in the two-dimensional
$x$-$y$ plane, a total of six parameters.
If one uses the Euclidean invariance of the $x$-$y$ plane
to take out translations and rotations,
one is left with three Euclidean-invariant initial conditions,
for example the object-string separation distance $r$,
the object speed
 \begin{equation}
 v \equiv |\dot{\mathbf r}_M|
    \equiv \sqrt{\dot{\mathbf r}_M\cdot\dot{\mathbf r}_M},
 \label{eq:36}
 \end{equation}
and the angle $\psi$ between
the string-object separation vector
${\mathbf r}$ and the object velocity vector 
 \begin{equation}
 {\mathbf v} \equiv \dot{{\mathbf r}}_M,
 \label{eq:37}
 \end{equation}
defined so that
 \begin{equation}
 {\mathbf r}\cdot{\mathbf v} = r v \cos{\psi}.
 \label{eq:38}
 \end{equation}
Instead of $\psi$, it is sometimes convenient
below to use
 \begin{equation}
 C \equiv \cos{\psi}
 = {{\mathbf r}\cdot{\mathbf v}\over r v}
 \label{eq:39a}
 \end{equation}
or
 \begin{equation}
 X \equiv \sin{\psi\over 2} = \sqrt{1-C \over 2}
 = \sqrt{{r v -{\mathbf r}\cdot{\mathbf v}\over 2 r v}}.
 \label{eq:39b}
 \end{equation}

	Note that the $v$ given by Eq. (\ref{eq:36})
as the speed of the object in the frame in which
the string was initially at rest,
and used with this meaning henceforth in this paper,
is not the same as the $v_r$ used in Section 2,
e.g., in Eq. (\ref{eq:15}),
which denoted $|\dot{{\mathbf r}}|$,
the instantaneous relative speed
of the object and the string,
and it is also not the same as the $v_s$ used
in Section 3, which by Eq. (\ref{eq:18})
denoted what we are now calling $|\dot{{\mathbf r}}_s|$,
the speed of the point on the string nearest the object,
in the frame in which the string was initially at rest.

	If one takes out the time-translation invariance,
one is left with only two initial conditions,
which can be the initial speed $v_0$ at $t = -\infty$
and the impact parameter $b = r\sin{\psi}$ at $t = -\infty$,
where $r$ is infinite and $\psi$ is $-\pi$.
There Eqs. (\ref{eq:35}) imply that $\dot{{\mathbf r}}_s = 0$,
and one can choose the initial string position to be
${\mathbf r}_s = 0$ (the string
initially at rest at $x_s = y_s = 0$)
and the object to be coming in from $x_M = -\infty$
with $y_M = b$ and with velocity $v_0$ in the positive
$x$-direction initially at $t = -\infty$.
These initial conditions
and the first of Eqs. (\ref{eq:35}) imply
that
 \begin{equation}
 {\mathbf r}_s = {\mathbf v}_0 - \dot{{\mathbf r}}_M
		= {\mathbf v}_0 - {\mathbf v}. 
 \label{eq:40}
 \end{equation}
Inserting this into the second of Eqs. (\ref{eq:35})
gives a single second-order differential equation
for the object vector position ${\mathbf r}_M(t)$,
although the second term in the rightmost
expression of Eqs. (\ref{eq:35})
makes this single equation
nonlinear in $\ddot{{\mathbf r}}_M$.

	As the object comes in from infinity,
it will be pulled toward the string,
so that at first it will speed up and bend downward,
and $y_M$ will decrease below its initial value of $b$.
Eventually the object will either be captured
by the string (so that $r$ and $v$ tend to zero),
or it scatter so that $r$ increases to infinity again,
depending on the initial velocity $v_0$
and the impact parameter $b$.
For each $v_0$, the boundary between
these two qualitatively different asymptotic
behaviors is given by the critical impact parameter
$b_{\rm crit}(v_0)$:
for $b < b_{\rm crit}(v_0)$ the object is captured,
and for $b > b_{\rm crit}(v_0)$ the object scatters
to $r = \infty$ with $v$ asymptotically greater than zero.
For $b = b_{\rm crit}(v_0)$, the marginal capture case,
$r$ reaches a minimum
and then slowly increases back to $\infty$,
but with $v$ asymptotically decreasing toward zero,
so that the object essentially has precisely
the escape velocity from the string.

	The main goal of this paper,
besides deriving the coupled equations of motion
(\ref{eq:31}) or (\ref{eq:35})
given above, is to calculate the
critical impact parameter $b_{\rm crit}(v_0)$
as a function of the initial velocity $v_0$
when this velocity is low compared with
the speed of light.
I shall derive one approximate formula
when $L_u/T_u \ll v_0 \ll c$,
the regime in which the string acts
essentially as a test string
at the critical impact parameter
(at least until $r$ reaches its minimum
value; during the late stages when
$r$ slowly increases back to $\infty$,
the back reaction of the string on the
object becomes important),
and another approximate formula
when $v_0 \ll L_u/T_u \ll c$,
the regime in which the string
acts stiffly and is only slightly
perturbed by the force from the object
(though this small perturbation
is essential for causing the object
to lose enough energy to become bound).

	In this paper I shall focus on the case in which
the ratio of the second term to the first term on the
extreme right hand side of Eqs. (\ref{eq:35}),
$4\mu r (\dot{{\mathbf r}}\cdot\dot{{\mathbf r}})$,
is much less than unity, so we shall omit
the last term of Eqs. (\ref{eq:35}).
For the dominant part of the scattering,
$r$ is of the order of magnitude of
the impact parameter $b$, and
$\dot{{\mathbf r}}\cdot\dot{{\mathbf r}}$
is of the order of magnitude of the initial
velocity $v_0$,
so this ratio is of the order of magnitude
of $\mu b v_0^2$ (in units with $T_u = L_u = 1$).
For $v_0 \ll L_u/T_u$,
we shall find below that
$b_{\rm crit} \propto v_0^{-7/5}$,
so for very small $v_0$,
certainly $\mu b v_0^2 \ll 1$
for the case of marginal capture.
For $v_0$ of order unity ($L_u/T_u$),
the truncated Eqs. (\ref{eq:35})
(with the last term omitted)
have no small or large parameters,
so we expect that then $b_{\rm crit}$
is also of order unity ($L_u$), giving
$\mu b_{\rm crit} v_0^2 \sim \mu \ll 1$.
For $v_0 \gg 1$,
$b_{\rm crit} \approx 2 v_0^{-1/2}$, so
$\mu b_{\rm crit} v_0^2 \sim \mu v_0^{3/2}$.
This is also small compared to unity when
$v_0 \ll \mu^{-2/3} \sim (1-Q^2/M^2)^{1/3} c$.

	Therefore, for calculating
the critical impact parameter
$b_{\rm crit}(v_0)$,
it appears to be a good approximation
to neglect the last term of Eqs. (\ref{eq:35})
for velocities low compared with
the speed of light times the cube root
of $(1-Q^2/M^2)$.
Unless the object is significantly charged
(e.g., nearly extremely charged
if the object is a black hole),
the limitation of considering
only velocities low compared with the speed
of light is sufficient for dropping
the last term of Eqs. (\ref{eq:35})
when one calculates $b_{\rm crit}(v_0)$.
It is apparently not sufficient when
$(1-Q^2/M^2)^{1/3} c \ll v_0 \ll c$,
but the calculation of $b_{\rm crit}(v_0)$
in this regime for nearly extremely charged objects
will be left as an exercise for the reader.

	When we drop the last term of
Eqs. (\ref{eq:35}), use Eq. (\ref{eq:40})
to solve the first part of these equations,
and define
 \begin{equation}
 {\mathbf R} = {\mathbf r}_M - {\mathbf v}_0,
 \label{eq:41}
 \end{equation}
then what remains of Eqs. (\ref{eq:35})
becomes the single second-order vector equation
 \begin{equation}
 \ddot{\mathbf R}
 = -{{\mathbf R}+\dot{{\mathbf R}}
 \over |{\mathbf R}+\dot{{\mathbf R}}|^3}.
 \label{eq:42}
 \end{equation}
In terms of ${\mathbf R}$, one has
 \begin{equation}
 {\mathbf r}_M = {\mathbf v}_0 + {\mathbf R},
 \label{eq:43}
 \end{equation}
 \begin{equation}
 {\mathbf r}_s = {\mathbf v}_0 - \dot{{\mathbf R}},
 \label{eq:44}
 \end{equation}
 \begin{equation}
 {\mathbf r} = {\mathbf R} + \dot{{\mathbf R}},
 \label{eq:45}
 \end{equation}
 \begin{equation}
 {\mathbf v} \equiv \dot{{\mathbf r}}_M = \dot{{\mathbf R}}.
 \label{eq:46}
 \end{equation}

	Alternatively, Eq. (\ref{eq:42})
can be written as two first-order vector equations,
 \begin{equation}
 \dot{{\mathbf r}} = {\mathbf v} - {{\mathbf r}\over r^3},
 \label{eq:47}
 \end{equation}
 \begin{equation}
 \dot{{\mathbf v}} =  - {{\mathbf r}\over r^3}.
 \label{eq:48}
 \end{equation}
If we instead go to the three Euclidean-invariant
quantities $r$, $v$, and either
$C = \cos{\psi} = {\mathbf r}\cdot{\mathbf v}/(r v)$
or $X = \sin{(\psi/2)} = \sqrt{(1-C)/2}$,
one gets the three first-order scalar equations
 \begin{equation}
 \dot{r} = C v - {1 \over r^2},
 \label{eq:49}
 \end{equation}
 \begin{equation}
 \dot{v} = - {C \over r^2},
 \label{eq:50}
 \end{equation}
and either
 \begin{equation}
 \dot{C} = (1-C^2)\left({v \over r}-{1\over r^2 v}\right)
 \label{eq:51}
 \end{equation}
or
 \begin{equation}
 \dot{X} = -X(1-X^2)\left({v \over r}-{1\over r^2 v}\right).
 \label{eq:52}
 \end{equation}

	Another choice of variables is motivated
by the fact that in the stiff-string regime,
where $r^2 v \gg 1$, the equations are approximately
those of a unit-mass Keplerian particle orbiting a stationary
string that exerts a unit inverse-square force on the particle,
$\dot{{\mathbf r}} \approx {\mathbf v}$
and $\dot{{\mathbf v}} =  - {\mathbf r}/r^3$.
For such a particle, conserved quantities are
the energy,
 \begin{equation}
 E \equiv {1\over 2}v^2 - {1\over r},
 \label{eq:53}
 \end{equation}
the angular momentum,
 \begin{equation}
 L \equiv r v \sin{\psi} = r v \sqrt{1-C^2} = 2 r v X \sqrt{1-X^2},
 \label{eq:54}
 \end{equation}
one-eighth the cube of the angular momentum,
 \begin{equation}
 h \equiv {1\over 8} L^3 = r^3 v^3 (X^2 - X^4)^{3/2},
 \label{eq:55}
 \end{equation}
the eccentricity of the orbit,
 \begin{equation}
 e \equiv \sqrt{1 + 2E L^2},
 \label{eq:56}
 \end{equation}
and one-fourth the excess of the square
of the eccentricity over unity,
 \begin{equation}
 g \equiv {1\over 4}(e^2 - 1) = {1\over 2}E L^2
 = (r^2 v^4 - 2 r v^2) X^2 (1-X^2)
 = [(r v^2 - 1)^2 - 1](X^2 - X^4).
 \label{eq:57}
 \end{equation}
Only two of these quantities are independent,
and to get fairly simple differential equations
below, I shall focus on $g$ and $h$.
In terms of them and $X^2 - X^4 = (1/4)\sin^2{\psi}$,
one may readily solve algebraically for $r$ and $v$:
 \begin{equation}
 r = {h^{2/3}\over
 \sqrt{X^2-X^4}(\sqrt{X^2-X^4} + \sqrt{X^2-X^4+g}\:)},
 \label{eq:58}
 \end{equation}
 \begin{equation}
 v = h^{-1/3}(\sqrt{X^2-X^4} + \sqrt{X^2-X^4+g}\:).
 \label{eq:59}
 \end{equation}

	Because the string is in actuality
not completely stiff but moves toward the object,
the quantities $E$, $L$, $h$, $e$, and $g$
are not precisely conserved
in the coupled object-string equations
but instead are all decreasing with time, at the rates
 \begin{equation}
 \dot{E} = - {1\over r^4},
 \label{eq:60}
 \end{equation}
 \begin{equation}
 \dot{L} = - {L\over r^3},
 \label{eq:61}
 \end{equation}
 \begin{equation}
 \dot{h} = - {3h\over r^3}
 = -3 h^{-1}(X^2-X^4)^{3/2}(\sqrt{X^2-X^4} + \sqrt{X^2-X^4+g}\:)^3,
 \label{eq:62}
 \end{equation}
 \begin{equation}
 \dot{e} = - {L^2\over e r^4}(r v^2 - 1) = -{v^3\over e}\dot{C},
 \label{eq:63}
 \end{equation}
 \begin{eqnarray}
 \dot{g}  &=& - {L^2\over 2 r^4}(r v^2 - 1)
 = -{1\over 2} v^3 \dot{C} = 2 v^3 X \dot{X}
	\nonumber \\
 &=& - 2 h^{-2}(X^2-X^4)^{3/2}\sqrt{X^2-X^4+g}\:
	(\sqrt{X^2-X^4} + \sqrt{X^2-X^4+g}\:)^4.
 \label{eq:64}
 \end{eqnarray}

	Now we can insert Eq. (\ref{eq:58}) and Eq. (\ref{eq:59})
into Eq. (\ref{eq:52}) to get
 \begin{equation}
 \dot{X} = -{(X^2-X^4)^{3/2}\over h X}\sqrt{X^2-X^4+g}\:
	(\sqrt{X^2-X^4} + \sqrt{X^2-X^4+g}\:).
 \label{eq:65}
 \end{equation}
Then dividing $\dot{g}$ and $\dot{h}$ by $\dot{X}$
gives two coupled first-order differential equations
for the coupled motion of the object and the string:
 \begin{equation}
 {dg \over dX}
 = {2\over h}X(\sqrt{X^2-X^4} + \sqrt{X^2-X^4+g}\:)^3,
 \label{eq:66}
 \end{equation}
 \begin{equation}
 {dh \over dX}
 = {3X(\sqrt{X^2-X^4} + \sqrt{X^2-X^4+g}\:)^2
	\over \sqrt{X^2-X^4+g}}.
 \label{eq:67}
 \end{equation}

	At $t=-\infty$, one has the object approaching
the string, and so the angle $\psi$ between
the object velocity ${\mathbf v}$ and the infinitely-long
string-object separation vector ${\mathbf r}$
is $\pi$, giving $X = \sin{(\psi/2)} = 1$ initially.
The energy is then $E_0 = v_0^2/2$,
and the angular momentum is then $L_0 = b v_0$, so
 \begin{equation}
 g(X=1) = {1\over 2}E_0 L_0^2 = {1\over 4} b^2 v_0^4,
 \label{eq:68}
 \end{equation}
 \begin{equation}
 h(X=1) = {1\over 8}L_0^3 = {1\over 8} b^3 v_0^3.
 \label{eq:69}
 \end{equation}

	Then as $X$ decreases, so do $g$ and $h$.
If the object scatters and remains unbound,
the energy and hence $g$ remain positive as
the object returns to infinitely large $r$,
but now with the object moving away from
the string, so ${\mathbf v}$ and ${\mathbf r}$
asymptotically approach the same direction,
giving $\psi = 0$ and hence $X = 0$.
If the object becomes bound to the string,
$E = v^2/2 - 1/r$ becomes negative, and hence also $g$.
Eq. (\ref{eq:67}) becomes singular where
$\sqrt{X^2-X^4+g}$ goes to zero,
but one can switch back to
Eqs. (\ref{eq:49})-(\ref{eq:51}) or (\ref{eq:52})
to continue the calculation,
as $\dot{X}$ and $\sqrt{X^2-X^4+g}$
reverse sign and the evolution continues,
with $X$ oscillating as the object spirals
in to join the string after a finite time.

	In the intermediate case in which
the impact parameter $b$ has the critical
value $b_{\rm crit}(v_0)$,
$E$ and $g$ tend to zero
as the object slowly moves to infinite
separation $r$ with essentially just
the escape velocity.
In this case, like the scattering case
in which $g$ stays positive,
the object velocity ${\mathbf v}$ and the
string-object separation vector ${\mathbf r}$
asymptotically become parallel,
giving $X=0$ at $t=\infty$.

	It is now easy to calculate
a one-parameter family of $v_0$'s
and their corresponding $b_{\rm crit}(v_0)$'s
using Eqs. (\ref{eq:66}) and (\ref{eq:67}),
without requiring a shooting method
from initial conditions at $t = -\infty$ or $X=1$.
Instead, one evolves backward in time
from $t=+\infty$ or $X=0$,
where one sets $g(X=0)=0$ but uses 
 \begin{equation}
 \eta \equiv h(X=0) = {1\over 8}L^3(t=\infty)
 \label{eq:70}
 \end{equation}
as the single free parameter for
determining the one-parameter
family of critical solutions.
Each solution is evolved to $X=1$,
where one evaluates $g(\eta,X=1)$ and $h(\eta,X=1)$
and then inverts Eqs. (\ref{eq:68}) and (\ref{eq:69})
to get
 \begin{equation}
 v_0(\eta) = g(\eta,X=1)^{1/2} h(\eta,X=1)^{-1/3},
 \label{eq:71}
 \end{equation}
 \begin{equation}
 b_{\rm crit}(v_0(\eta))
 = 2 g(\eta,X=1)^{-1/2} h(\eta,X=1)^{2/3}.
 \label{eq:72}
 \end{equation}

\section{The critical impact parameter
in the stiff-string regime, $v_0 \ll L_u/T_u$}

\hspace{.20in}	Now we shall find approximate
analytic solutions of the evolution
Eqs. (\ref{eq:66}) and (\ref{eq:67})
in the marginally bound case
($g(X=0) = 0$ or $E(t=\infty) = 0$)
in the two limiting cases in which
$\eta = h(X=0) = L^3(t=\infty)/8$
is either very large or very small.

	For $\eta \gg 1$, the factor of $h$
in the denominator of Eq. (\ref{eq:66})
means that to lowest order $g(X)$
stays very small and can be neglected
on the right hand sides of
Eqs. (\ref{eq:66}) and (\ref{eq:67}).
Then one gets
 \begin{eqnarray}
 g(X) &\approx&
 {1\over 64\eta}(12\psi - 8\sin{2\psi}+\sin{4\psi})
	\nonumber \\
 &=& {1\over 8\eta}
	[3\sin^{-1}{X}+(-3-2X^2+24X^4-16X^6)\sqrt{X^2-X^4}],
 \label{eq:73}
 \end{eqnarray}
 \begin{equation}
 h(X) \approx \eta + {3\over 8} (2\psi - \sin{2\psi})
 = \eta + {3\over 2}[\sin^{-1}{X}+(-1+2X^2)\sqrt{X^2-X^4}].
 \label{eq:74}
 \end{equation}
This is sufficiently accurate to give 
 \begin{equation}
 h(\eta,X=1)
 = \eta\left[1 + {3\pi\over 4\eta} + O({1\over \eta^2})\right].
 \label{eq:75}
 \end{equation}
However, to get $g(\eta,X=1)$ with a relative error of only
$O(\eta^{-2})$ rather than $O(\eta^{-1})$,
one needs to insert
the approximate expressions (\ref{eq:73}) and (\ref{eq:74})
into the right hand side of Eq. (\ref{eq:66})
and evaluate it through the next-to-lowest order
in $\eta^{-1}$.  This gives
 \begin{equation}
 g(\eta,X=1)
 = {3\pi\over 16\eta}
	\left[1 + {0\over \eta} + O({1\over \eta^2})\right].
 \label{eq:76}
 \end{equation}

	If we insert these expressions into
Eqs. (\ref{eq:71}) and (\ref{eq:72})
to get $v_0(\eta)$ and $b_{\rm crit}(\eta)$
and then eliminate $\eta$ to get $b_{\rm crit}(v_0)$,
we find that
 \begin{equation}
 b_{\rm crit} = (6\pi)^{1/5} v_0^{-7/5}
 + {1\over 5}(6\pi)^{3/5} v_0^{-1/5} +O(v_0).
 \label{eq:77}
 \end{equation}
Large $\eta$ gives small
$v_0 \approx \sqrt{3\pi}/(4\eta^{5/6})$,
so Eq. (\ref{eq:77}) applies for
$v_0 \ll L_u/T_u \equiv 1$.
At the critical impact parameter,
this is the stiff-string regime,
where the string responds much less
than the object to the force between them.
If one restores the units $T_u$ and $L_u$
given by Eqs. (\ref{eq:32}) and (\ref{eq:33}),
one obtains the formula given in the Abstract
above for the stiff-string regime.

	One can give an order-of-magnitude
derivation of the critical impact parameter
to show that it indeed goes as
$v_0^{-7/5}$ for $v_0 \ll 1$.
What happens is that the object
comes in on a slightly hyperbolic,
but nearly parabolic, orbit
with initial energy $E_0 = v_0^2/2$
just slightly positive.
Then as the object swings around
the string, the small motion of the string
toward the object drains just enough energy
from the object that it goes back out
on an asymptotically parabolic orbit
with $E$ approaching zero.

	In this analysis, use $\approx$
for approximate numerical equality under the
limits given (i.e., $x \ll 1$ implies that $1+x \approx 1$),
and use $\sim$ for order-of-magnitude equality
that has the correct powers of small quantities
(e.g., $2 x \sim x$).

	With impact parameter $b \gg 1/v_0$,
the angular momentum $L$
stays nearly constant at its initial value
$L_0 = b v_0 \gg 1$
(as I shall show momentarily),
so at the perispaggon,
the point on the object orbit nearest the string,
the minimal separation $r_m$ and the
object speed $v_m$ are related
by $r_m v_m \approx L \approx L_0 = b v_0$.
Assuming that $E = v^2/2 - 1/r \ll 1/b$
(which for $E \sim E_0$ requires $b \ll 1/v_0^2$,
so we need $1/v_0 \ll b \ll 1/v_0^2$ and hence
$v_0 \ll 1$), we get 
$v_m^2/2 \approx 1/r_m \gg v_0^2/2$.
Combining this with $r_m v_m \approx L_0$
then gives
 \begin{equation}
 r_m \approx {1\over 2}L_0^2,
 \label{eq:77b}
 \end{equation}
 \begin{equation}
 v_m \approx {2\over L_0}.
 \label{eq:77c}
 \end{equation}

	The time spent near perispaggon
is $t_m \sim r_m/v_m \sim L_0^3$.
During this time, the angular momentum
decreases at the rate given by Eqs. (\ref{eq:61}),
$\dot{L} = - L/r^3 \sim -L_0/r_m^3 \sim -1/L_0^5$,
and thus it decreases by a total amount
roughly $t_m$ as large, or $\Delta L \sim - 1/L_0^2$.
The magnitude of this decrease is much less than
$L_0$ for $L_0 \gg 1$ or $b \gg 1/v_0$, as claimed
above, so in this regime the angular momentum
is approximately constant.

	The energy decreases at the rate
given by Eqs. (\ref{eq:70}),
$\dot{E} = -1/r^4 \sim -1/r_m^4 \sim 1/L_0^8$,
and multiplying this by the time $t_m$
gives a total decrease
$-\Delta E \sim 1/L_0^5 \approx 1/(b v_0)^5$.
For the final energy to be zero,
this decrease needs to equal the initial
energy, giving
$v_0^2 \sim E_0 \sim -\Delta E \sim 1/(b v_0)^5$
or $b = b_{\rm crit} \sim v_0^{-7/5}$.
For $v_0 \ll 1$, this is indeed within
the allowed range
$1/v_0 \ll b_{\rm crit} \sim v_0^{-7/5} \ll 1/v_0^2$
of the approximations used above.

	One can easily get the correct coefficient
$(6\pi)^{1/5}$ of the $v_0^{-7/5}$ term
for $b_{\rm crit}$
by calculating the time integral of the
energy loss rate $\dot{E} = -1/r^4$
for a parabolic orbit of angular momentum
$L_0 = b v_0$,
but that confirmation of the precise form
of the first term of the right hand side of
Eq. (\ref{eq:77}) will be left as an exercise
for the reader.
Similarly, one should be able to get the
second term on the right hand side of
Eq. (\ref{eq:77}) from first-order perturbations
due to the motion of the string
and the resultant loss of the energy and
angular momentum of the object.

\section{The critical impact parameter
in the test-string regime, $v_0 \gg L_u/T_u$}

\hspace{.20in}	Now I shall consider the opposite case,
$\eta \equiv h(X=0) = L^3(t=\infty)/8 \ll 1$,
which leads to $v_0 \gg L_u/T_u \equiv 1$
and $b_{\rm crit} \ll L_u \equiv 1$.
This is the test-string limit,
for which the lowest-order approximation
is to ignore the string tension
and the effect of the string on the object.
The critical impact parameter
in this limit was found in \cite{Page}
to be $b_{\rm crit} \simeq 2 v_0^{-1/2}$
after converting to the units used
in this paper.
Here I shall rederive this result
and give the lowest-order
correction in the small quantity $1/v_0$.

	In this test-string regime $v_0 \gg 1$,
the qualitative picture is that initially
the object moves with very nearly
constant velocity, and the string gets bent
toward the object as described in \cite{Page}.
In that paper, where the back reaction of the
string on the object was totally ignored,
the final situation, when the impact parameter
$b$ has the critical value $b_{\rm crit}$
for marginal capture,
is for the string to develop
a kink that asymptotically trails the object
at a distance of $v_0^{-1/2} = b_{\rm crit}/2$
behind it, an unstable equilibrium point
where the force of the object on the string
causes the string to follow the object
at precisely the same speed as the object,
by Eq. (\ref{eq:29}).
In the complete test-string approximation,
the deceleration of the object
is ignored, so it continues moving forever,
pulling an ever longer kink of string
behind it.

	When one does include the back reaction
effect of the string tension on the object
but stays in the test-string regime of $v_0 \gg 1$,
the picture during the first stage of the evolution
in the marginal capture case
is very similar to that of the complete
test-string approximation used in \cite{Page},
so the position on the string nearest the object
approaches quite near to what would be the
unstable equilibrium separation behind the object
if the object continued moving with constant velocity.
But then in actuality the object slows down
as the string weakly pulls back on it
(somewhat analogous to the way a fish slows down when
attached to a weak fishing line \cite{CarterPrivate}).
If it had been at the unstable equilibrium separation
for the velocity of the object
(i.e., the separation that would be the equilibrium one
if the object continued moving with constant velocity),
then the slowing down of the object would
eventually lead to its falling into the string.
For the object to be just marginally
not falling into the string,
the separation must be greater,
by a precise tiny amount, than
what would be the equilibrium separation
at constant velocity.
At this slightly greater separation, the force
on the kink of the string trailing the object
would be just slightly less than needed
to move it at the speed of the object,
so the string moves slightly slower than the object
and hence trails behind it
at an indefinitely increasing separation,
gradually slowing down more and more.

	If the object had had an impact parameter $b$
slightly greater than $b_{\rm crit}(v_0)$,
it would maintain a velocity bounded away from
zero as its separation from the string
increased indefinitely.
On the other hand, if the object
had $b$ slightly smaller than $b_{\rm crit}(v_0)$,
the object would eventually
slow down more than the string,
so that the separation would stop increasing
and would henceforth decrease to zero
as the object falls into the string.
But in the marginal capture case
with impact parameter $b = b_{\rm crit}(v_0)$,
the separation $r$ between the string
and the object, although continuing
to grow indefinitely, grows slowly
enough that the object, as well as the string,
continues to slow down just fast enough
that it asymptotically approaches,
but never reaches, zero velocity.

	We can put this discussion into a more quantitative
form in the following way:  For $g \ll X^2 \ll 1$,
Eqs. (\ref{eq:66}) and (\ref{eq:67}) become
 \begin{equation}
 {dg \over dX}
 = {16 X^4 + 12X^2 g + O(X^6) + O(g^2) \over h},
 \label{eq:78}
 \end{equation}
 \begin{equation}
 {dh \over dX} = 12 X^2 + O(X^4) + O(g^2),
 \label{eq:79}
 \end{equation}
with the critical solutions ($g=0$ and $h=\eta > 0$ at $X=0$)
 \begin{equation}
 g(X)
 = {16 \over 5}\left({X^5 \over \eta} - {X^8 \over \eta^2}\right)
 + O\left({X^7 \over \eta}\right)
 + O\left({X^{10} \over \eta^2}\right)
 + O\left({X^{11} \over \eta^3}\right),
 \label{eq:80}
 \end{equation}
 \begin{equation}
 h(X) = \eta + 4X^3 + O(X^5) + O(\eta^{-1}).
 \label{eq:81}
 \end{equation}

	To get the time dependence of these solutions,
we can use Eq. (\ref{eq:65}),
which for $g \ll X^2 \ll 1$ becomes
 \begin{equation}
 \dot{X}
 = {4 X^4 + 3X^2 g + O(X^6) + O(g^2) \over 2h}.
 \label{eq:82}
 \end{equation}
Inserting $g(X)$ and $h(X)$ from
Eqs. (\ref{eq:80}) and (\ref{eq:81})
into this and integrating gives
 \begin{equation}
 t = {\eta \over 6X^3} - {4\over 5}\ln{X}
 + O\left({X^3 \over \eta}\right),
 \label{eq:83}
 \end{equation}
dropping the irrelevant constant of integration.
For large $t$ this can be inverted to give
 \begin{equation}
 X = \left({\eta\over 6t}\right)^{1/3}
	\left[1 + {4\over 45t}\ln{6t\over\eta}
	 + O\left({\ln^2{t} \over t^2}\right)\right],
 \label{eq:84}
 \end{equation}
Then from Eqs. (\ref{eq:80}), (\ref{eq:81}),
(\ref{eq:58}), and (\ref{eq:59}), one may get
 \begin{equation}
 g(t) \approx
	{16\over 5}\eta^{2/3}(6t)^{-5/3}
	\left(1 + {4\over 9t}\ln{6t\over\eta}\right),
 \label{eq:85}
 \end{equation}
 \begin{equation}
 h(t) \approx
	\eta \left(1 + {3\over 2t}
	 + {8 \over 45t^2}\ln{6t\over\eta}\right),
 \label{eq:86}
 \end{equation}
 \begin{equation}
 r(t) \approx
	{1\over 2}(6t)^{2/3}
	 \left(1 - {8 \over 45t}\ln{{6t\over\eta}}\right),
 \label{eq:87}
 \end{equation}
 \begin{equation}
 v(t) \approx
	2 (6t)^{-1/3}
	 \left(1 + {4 \over 45t}\ln{6t\over\eta}\right).
 \label{eq:88}
 \end{equation}

	Alternatively, one can divide
Eq. (\ref{eq:49}) by Eq. (\ref{eq:50}) to get
 \begin{equation}
 {dr\over dv} = \sec{\psi} - r^2 v
	 \approx 1 - r^2 v
 \label{eq:89}
 \end{equation}
for $X \equiv \sin{\psi/2} \ll 1$,
where one may recall that $\psi$
is the angle between the velocity
of the object and the string-object
separation, so $\psi$ goes from $\pi$ initially
(as the object heads toward the string)
to $0$ finally for marginal capture
(as the object heads away from the string
with slower and slower velocity).
When $X$ can be neglected,
Eq. (\ref{eq:89}) leads to the series solution
 \begin{equation}
 {1\over r} \approx
 {1\over 2}v^2-{1\over 20}v^5+{1\over 160}v^8
	-{7\over 8800}v^{11}+{1\over 9856}v^{14}+O(v^{17})
 \label{eq:90}
 \end{equation}
or
 \begin{equation}
 v(r) \approx
 \sqrt{2}\,r^{-1/2}+{1\over 5}r^{-2}+{\sqrt{2}\over 25}r^{-7/2}
 +{24\over 1375}r^{-5} + O(r^{-13/2}).
 \label{eq:91}
 \end{equation}

	All of the series expansions
given so far in this section
are good only when $t \gg 1$, $r \gg 1$,
and $v \ll 1$, in addition to $X \ll 1$
(i.e., when the object is moving very slowly
nearly directly away from the string
at very late times).
When $\eta \ll 1$, as one integrates
backward in time from the final condition
$g=0$ and $h=\eta$ at $X=0$
(where $t=\infty$, $r=\infty$, and $v=0$),
the approximate second equality of Eq. (\ref{eq:89})
remains valid so long as $X \ll 1$ and gives,
when $v$ becomes very large
and $r$ becomes very small,
 \begin{equation}
 {1\over r} \approx
 v^{1/2} - {1\over 4}v^{-1} - {5\over 32}v^{-5/2}
 - {15\over 64}v^{-4} - {1105\over 2048}v^{-11/2} + O(v^{-7}),
 \label{eq:92}
 \end{equation}
up to terms that decrease roughly
as the exponential of $-(4/3)v^{3/2}$,
so long as $X$ remains negligible.
Inverting this gives
 \begin{equation}
 v(r) \approx
 {1\over r^2}+{1\over 2}r+{1\over 8}r^4
 +{5\over 32}r^7 + O(r^{10})
 \label{eq:93}
 \end{equation}
for small $r$, again neglecting
all contributions from nonzero $X$
and terms exponential in $-4/(3r^3)$.

	To get the $X$-dependence of the
critical solution when $X \ll 1$
and to prepare the way for the corrections
when $X$ is no longer small,
divide Eq. (\ref{eq:52}) by Eq. (\ref{eq:50}) to get
 \begin{equation}
 {dX\over dv} = {X(1-X^2)\over 1-2X^2}
	 \left(1-{1\over r v^2}\right) r v
	\approx X \left(1-{1\over r v^2}\right) r v,
 \label{eq:94}
 \end{equation}
with the approximate equality applying when $X \ll 1$.

	When $v \ll 1$ so that Eq. (\ref{eq:90}) is applicable,
Eq. (\ref{eq:94}) becomes
 \begin{equation}
 {dX\over dv} \approx {X \over v},
 \label{eq:95}
 \end{equation}
which, when matched to the boundary condition
that $L = 2 r v X \sqrt{1-X^2}$ is equal to $2\eta^{1/3}$
at $X=0$, gives
 \begin{equation}
 X \approx {1\over 2} \eta^{1/3} v
 \approx {\eta^{1/3} \over \sqrt{2r}}
 \label{eq:96}
 \end{equation}
for $v \ll 1$, $r \gg 1$, or $X \ll \eta^{1/3}$.

	At $X \sim \eta^{1/3} \ll 1$,
$v$ surpasses unity and thereafter becomes large,
and $r$ becomes small
(until $X$ gets near 1 and $r$ becomes large again
in this evolution backward in time).
When $v \gg 1$
so that Eq. (\ref{eq:92}) is applicable,
with $r v^2 \approx v^{3/2} \gg 1$,
Eq. (\ref{eq:94}) becomes
 \begin{equation}
 {dX\over dv} \approx {X(1-X^2) r v\over 1-2X^2}
	\approx X r v,
 \label{eq:97}
 \end{equation}
where the first approximate equality
applies for all $X \gg \eta^{1/3}$ and
the second approximate equality
applies for $\eta^{1/3} \ll X \ll 1$.

	Eq. (\ref{eq:97}), along with Eq. (\ref{eq:92}),
implies that once $v$ gets large,
$X$ grows roughly exponentially with $(2/3)v^{3/2}$.
After $X$ passes $v^{-1/2}$, $dv/dX$ becomes small,
and the large $v$ changes very little during the rest
of the backward evolution to the initial conditions
at $X=0$ ($t=-\infty$, $v=v_0$).
Since this exponential growth of $X$ with $v$
starts at $X \sim \eta^{1/3}$ where $v \sim 1$
and ends at $X \sim 1$ where $v \sim v_0$,
one gets
 \begin{equation}
 v_0 \sim (-{1\over 2}\ln{\eta})^{2/3}
 \label{eq:98}
 \end{equation}
for very small $\eta$.
This very slow growth of $v_0$ with $1/\eta$
implies that when one solves
Eqs. (\ref{eq:66}) and (\ref{eq:67})
numerically with the numerical initial conditions
$g(X=0) = 0$ and $h(X=0) = \eta$
(final conditions in time for marginal capture),
one must choose $\eta$ to be extremely
tiny to get a large value of $v_0$,
and for the approximations of this section,
one must assume by $\eta \ll 1$ not just that
$\eta$ is more than an order of magnitude (say)
smaller than unity, but that $-\ln{\eta}$
is more than an order of magnitude (say)
larger than unity.

	Once the marginal capture solution
for $\eta \ll 1$ has been carried past
$X \sim (-\ln{\eta})^{-1/3}$
where $dv/dX$ becomes small,
$v$ stays nearly constant and one is
in the test-string regime treated in \cite{Page},
where one can to first approximation
neglect the back reaction of the string
on the object (the object's acceleration
toward the string).  In this paper, however,
we want to treat this back reaction
perturbatively in this test-string regime
and get the first-order correction
to the test-string critical impact parameter
formula $b_{\rm crit} \simeq 2 v_0^{-1/2}$
given in \cite{Page}.

	For this purpose it helps to change
the dependent variables of integration
from the $g(X)$ and $h(X)$ used in
Eqs. (\ref{eq:66}) and (\ref{eq:67})
to two other variables that would
be constant if the test-string
approximation were completely accurate.
One such variable is $v$, the magnitude
of the velocity of the object
(in the frame in which the string
was initially at rest along the $z$-axis).
A second such variable is motivated
by the test-string Eq. (67) of \cite{Page},
where the polar coordinate angle $\theta$
defined just before Eq. (54) of that paper
is essentially the same as $\psi$ here,
where $r$ there is the same as $r$ here,
where $z$ given there just before
Eq. (54) as $r \sin{\theta}$
is hence the same as $r \sin{\psi} = 2 r X \sqrt{1-X^2}$
in the notation of the present paper,
and where the constant $L$ given by Eq. (50)
(not to be confused with the angular momentum $L$
of the present paper)
is the same as $v_0^{-1/2}$
in our present units $T_u \equiv 1$ and $L_u \equiv 1$.
Then Eq. (67) of \cite{Page}, plus the equation
$b_{\rm crit} \simeq 2L$ three lines below, imply that
 \begin{equation}
 b^2 -  b_{\rm crit}^2 \simeq 4 r^2 X^2 (1-X^2) - 4 v_0^{-1} X^2
 \label{eq:99}
 \end{equation}
is a constant of motion in the test-string approximation,
and furthermore that it is zero for marginal capture.
If we use the fact that $v \simeq v_0$ to replace
$v_0$ by $v$ and then multiply the expression on the
right hand side by $v/4$, we get the quantity
 \begin{equation}
 w \equiv r^2 v (X^2-X^4) - X^2 = {L^2\over 4 v} - X^2
 = {h \over \sqrt{X^2-X^4}+\sqrt{X^2-X^4+g}} - X^2.
 \label{eq:100}
 \end{equation}

	Eqs. (\ref{eq:66}) and (\ref{eq:67}) now imply that
the alternative dependent variables $v$ and $w$
obey the differential equations
 \begin{equation}
 {dv \over dX}
 = {X(1-2X^2)\over \sqrt{v(X^2-X^4)(X^2+w)}}
 \left[1-\sqrt{{X^2-X^4\over v^3(X^2+w)}}\right]^{-1},
 \label{eq:101}
 \end{equation}
 \begin{equation}
 {dw \over dX}
 = {X[X^2-(1-2X^2)w]\over\sqrt{v^3(X^2-X^4)(X^2+w)}}
 \left[1-\sqrt{{X^2-X^4\over v^3(X^2+w)}}\right]^{-1}.
 \label{eq:102}
 \end{equation}

	We can see from Eq. (\ref{eq:100}) that $w$
is not a convenient variable to integrate all the way
from $X=0$, since sufficiently near there that
$h \approx \eta$ and $g \ll 1$,
the last expression of Eqs. (\ref{eq:100}) implies that
we have $w \approx h/(2X)$, which diverges at $X=0$.
Alternatively, in the second-to-last expression
of Eq. (\ref{eq:100}), one gets a divergence
from the fact that $L$ goes to the nonzero constant
$2\eta^{1/3}$ at $X=0$, but $v$ goes to zero there.
However, from Eq. (\ref{eq:92}) for $\eta^{1/3} \ll X \ll 1$,
where $v \gg 1$ and $r \approx v^{-1/2} \ll 1$,
one can see that then $w \approx (1/2)v^{-3/2}X^2$,
which is negligibly small for small enough $X$.
E.g., for $X \sim \eta^{1/3} \sim e^{-(2/3)v_0^{3/2}}$,
where $v \sim 1$ and $r \sim 1$,
$w \sim e^{-(2/3)v_0^{3/2}}$
is exponentially small for large $v_0$.
Therefore, with only an exponentially small error,
we can set $w=0$ and $v \gg 1$ as initial conditions
in integrating Eqs. (\ref{eq:101}) and (\ref{eq:102})
from some very small $X$, such as $\eta^{1/3}$,
which for practical purposes we can replace with 0
with only an additional exponentially small error.

	Because Eqs. (\ref{eq:101}) and (\ref{eq:102})
have $v^{1/2} \gg 1$ and $v^{3/2} \gg 1$, respectively,
in the denominators of the right hand sides,
whereas the other factors are of order unity
(except for harmless square-root singularities
in $1-X^2$ at $X=1$),
$v$ and $w$ change by only small amounts
as one integrates to $X=1$ to get the temporally initial
conditions for marginal capture.
To lowest order in $1/v_0$, one can
thus neglect $w$ on the right hand sides and rewrite
Eqs. (\ref{eq:101}) and (\ref{eq:102}) as
 \begin{equation}
 {dv \over dX}
 \approx {1-2X^2\over \sqrt{v(X^2-X^4)}}
 \approx {1-2X^2\over \sqrt{v_0(X^2-X^4)}},
 \label{eq:103}
 \end{equation}
 \begin{equation}
 {dw \over dX}
 \approx {X^2\over\sqrt{v^3(X^2-X^4)}}
 \approx {X^2\over\sqrt{v_0^3(X^2-X^4)}}.
 \label{eq:104}
 \end{equation}
Then Eq. (\ref{eq:104}) integrates to give
 \begin{equation}
 w(X=1) = {L_0\over 4v_0} - 1
 = {1\over 4} b_{\rm crit}^2 v_0 - 1
 = v_0^{-3/2} + O(v_0^{-3}).
 \label{eq:105}
 \end{equation}

From this we can readily solve for the
critical impact parameter,
in the units of $T_u$ and of $L_u$
given by Eqs. (\ref{eq:32}) and (\ref{eq:33}),
for large $v_0 \gg L_u/T_u \equiv 1$:
 \begin{equation}
 b_{\rm crit}(v_0) = 2 v_0^{-1/2} + v_0^{-2} + O(v_0^{-7/2}).
 \label{eq:106}
 \end{equation}
When the units are restored,
one gets the result given in the Abstract.

	An order-of-magnitude argument
for the $v_0^{-1/2}$ power law
of the leading term was given in \cite{Page}.
Summarizing it very briefly in the units used here,
if the object is incident with velocity $v_0 \gg 1$
and impact parameter $b$ such that it is not captured,
the distance of closest approach will be $r \sim b$,
the velocity at that point will be $v \sim v_0$,
the time when the string is roughly that close
will be $\sim r/v \sim b/v_0$,
the length of string that bends significantly
will be the speed of light times this or $\sim cb/v_0$,
the bending angle will be $\sim 1/(cr^2) \sim 1/(cb^2)$,
and hence the total transverse bending of the string
will be the length of string significantly bent
times this bending angle, or $\sim 1/(b v_0)$.
Marginal capture requires that this total transverse
bending be of the same order as the impact parameter,
giving $b = b_{\rm crit} \sim 1/\sqrt{v_0}$.

\section{The critical impact parameter
over the entire range of nonrelativistic velocities}

\hspace{.20in}	Now that we have found analytically
Eq. (\ref{eq:77}) for the critical impact parameter
at very low initial velocities, $v_0 \ll L_u/T_u \ll c$,
and Eq.  (\ref{eq:106}) for $b_{\rm crit}(v_0)$
at high (but still nonrelativistic) velocities,
$L_u/T_u \ll v_0 \ll c$,
we would like to find it numerically
for all nonrelativistic velocities,
and ideally find a fairly simple formula
that gives a good fit to $b_{\rm crit}(v_0)$
for all $v_0$.

	It was straightforward to program
Mathematica 2.0 on my ancient NeXT
to integrate Eqs. (\ref{eq:66}) and (\ref{eq:67})
numerically from the temporally final conditions
$g=0$ and $h=\eta$ at $X=0$
back to the temporally initial conditions at $X=0$
and then use Eqs. (\ref{eq:71}) and (\ref{eq:72})
to solve for $v_0$ and $b_{\rm crit}(v_0)$.
It is slightly awkward that by this method
one cannot straightforwardly choose
the values of $v_0$ for which one gets
$b_{\rm crit}(v_0)$, but this is a small price
to pay for not having to use a shooting method
to find the critical or marginally bound solutions.
(One could use a shooting method to get
any particular $v_0$ desired,
but for determining the characteristics
of the function $b_{\rm crit}(v_0)$,
this is hardly necessary.)

	Another slight awkwardness is that to get large
values of the initial velocity $v_0$,
one must use exponentially small values
of $\eta$, and this can lead to some
numerical inaccuracy, but this was not
a serious problem for at least moderately large
values of $v_0$.

	I calculated over 50 pairs of
$(v_0, b_{\rm crit}(v_0))$,
ranging from $(0.00104198, 26910)$
through (for example) $(0.834385, 3.48671)$
to $(357.89, 0.105727)$.
Larger values of $v_0$
require such extraordinarily small values of $\eta$
that it would be somewhat difficult
by my numerical method to obtain them,
but what I could easily attain extends
well into the domain of validity of both
Eq. (\ref{eq:77}) and Eq. (\ref{eq:106}).

	The results from the numerical integrations
fit extremely well to the truncated series expansions
of Eq. (\ref{eq:77}) at very small $v_0$ and
of Eq. (\ref{eq:106}) at large $v_0$,
confirming both the two coefficients and the two exponents
of each of the two expressions (eight parameters
calculated analytically and confirmed numerically).

	One can combine Eqs. (\ref{eq:77}) and (\ref{eq:106})
into the interpolation formula
 \begin{equation}
 b_{\rm crit}(v_0) \simeq B_{(\alpha,\beta)}(v_0) \equiv
 \left[\left({6\pi\over v_0^7}\right)^{\alpha\over 5}
  + \left({4\over v_0}\right)^{\alpha\over 2}\right]
	^{1\over\alpha}
 + \left[\left({5^5 v_0\over 6^3 \pi^3}\right)^{\beta\over 5}
  + v_0^{2\beta}\right]^{-{1\over\beta}},
 \label{eq:107}
 \end{equation}
which depends not only on the eight parameters
(four coefficients and four exponents)
calculated analytically and given in
Eqs. (\ref{eq:77}) and (\ref{eq:106}),
but also on two new parameters,
$\alpha$ and $\beta$.
Let us see how well this formula works
for suitable $\alpha$ and $\beta$.

	In order that the first two terms
of the right hand side of Eq. (\ref{eq:77})
be the dominant terms of the right hand side
of Eqs. (\ref{eq:107}) at $v_0 \ll 1$,
one needs $\alpha > 4/3$ and $\beta > 0$,
and in order that the first two terms
of the right hand side of Eq. (\ref{eq:106})
be the dominant terms of the right hand side
of Eqs. (\ref{eq:107}) at $v_0 \gg 1$,
one needs $\alpha > 5/3$ and $\beta > 0$.
The combination of these two pairs of inequalities
gives what might be called Criterion A,
$\alpha > 5/3$ and $\beta > 0$.
If one wants to avoid any deviations
larger than the $O(v_0)$ term of Eq. (\ref{eq:77})
at small $V_0$,
one needs $\alpha\geq 8/3$ and $\beta\geq 2/3$,
and if one wants to avoid any deviations
larger than the $O(v_0^{-7/2})$ term of Eq. (\ref{eq:106})
at large $v_0$,
one needs $\alpha\geq 10/3$ and $\beta\geq 5/6$.
The combination of these latter two pairs of inequalities
gives what might be called Criterion B,
$\alpha\geq 10/3$ and $\beta\geq 5/6$.

	If one wants a simple choice of
the pair of fitting parameters $(\alpha,\beta)$
(e.g., a pair of integers),
Criterion A motivated my first choice,
$(\alpha,\beta) = (2,1)$.
As a crude fit, this is not too bad,
giving a maximum relative error
of slightly over 10\%, e.g. 10.382\% error at
$(v_0, b_{\rm crit}(v_0)) = (0.940172, 3.10142)$.
My second choice,
$(\alpha,\beta) = (3,1)$,
is the smallest pair of integers
that (almost) satisfies Criterion B.
This gives
 \begin{equation}
 b_{\rm crit}(v_0) \simeq B_{(3,1)}(v_0) \equiv
 \left[\left(6\pi v_0^{-7}\right)^{3/5}
  + 8 v_0^{-3/2}\right]^{1/3}
 + \left[\left({3125 v_0\over 216 \pi^3}\right)^{1/5}
  + v_0^2\right]^{-1}.
 \label{eq:108}
 \end{equation}
This approximate formula gives a much better bit,
with a maximum relative error of
only about 0.85\%, near
$(v_0, b_{\rm crit}(v_0)) = (1.79814, 1.78528)$.
(The position of the maximum relative error
was not calculated to the six-digit
figures given for $(v_0, b_{\rm crit}(v_0))$
here, but they are given to six digits so
that one can see how accurate
this, or some other, approximation
is at this sample point,
a point where the relative error
was about as large as I could find given the
six-digit apparent accuracy
of the Mathematica calculation.) 

	I then used the FindMinimum function
of Mathematica to minimize,
as a function of the pair $(\alpha, \beta)$,
the sum of the
100th powers of the relative errors of
$B_{(\alpha,\beta)}(v_0)$ over the first
45 pairs of values of
$(v_0, b_{\rm crit}(v_0))$
that I had calculated.
This sum is an analytic function
of $(\alpha, \beta)$
whose minimum should be at a
value of $(\alpha, \beta)$
which is near the point
where the maximum absolute value of the relative
error for the 45 points is minimized,
since the sum of the 100th power of the 
relatives errors should be dominated by
the largest relative error.

	Mathematica gave the minimization as occurring at
$(\alpha, \beta) = (4.00448, 1.36884)$,
though the result would no doubt have
been slightly different if I had used
a different set of $(v_0, b_{\rm crit}(v_0))$
and a different power of the errors.
Ideally one should use an infinite set of points
and an infinite power of the errors,
but I considered 45 points and a power of
100 as being high enough to give a good
feel for the true minimum, over $(\alpha, \beta)$,
of the maximum of the magnitude of the relative error.
The largest relative error I found for
this value of $(\alpha, \beta)$ was 0.427328\%.

	By comparison, minimizing the sum of the
squares (2nd powers) of the relative errors
led to $(\alpha, \beta) = (3.98412, 1.37903)$,
which is not that much different even for
this relatively small value of the power,
though the maximum error I found
for this choice of $(\alpha, \beta)$,
0.47019\%, is about 10\% higher.

	A fairly simple rational approximation
for the $(\alpha, \beta)$ that minimizes the maximum
relative error of $B_{(\alpha,\beta)}(v_0)$
is $(4, 15/11)$, which then leads to
 \begin{equation}
 b_{\rm crit}(v_0) \simeq B_{(4,15/11)}(v_0) \equiv
  \left[\left({6\pi\over v_0^7}\right)^{4\over 5}
  + \left({2\over \sqrt{v_0}}\right)^4\right]
	^{1\over 4}
 + \left[\left({3125 v_0\over 216 \pi^3}\right)^{3\over 11}
  + v_0^{30\over 11}\right]^{-{11\over 15}}.
 \label{eq:109}
 \end{equation}
Here the number 4 in the exponents of the first term
corresponds to $\alpha$,
and the number 11 in the exponents of the second term
corresponds to $15/\beta$,
so if one wanted different values of
$(\alpha,\beta)$, one could simply change
these two numbers in this expression.
This approximation gives a maximum relative
error of just under 0.43\%
(I found a maximum of 0.429991\%), near
$(v_0, b_{\rm crit}(v_0)) = (0.820853, 3.54436)$.
Thus this $B_{(4,15/11)}(v_0)$
is over 99\% as good as $B_{(4.00448,1.36884)}(v_0)$
(has a maximum relative error less than 1.01 times
the maximum relative error of $B_{(4.00448,1.36884)}(v_0)$)
and is roughly twice as good an approximation
(i.e., with roughly half the maximum relative error)
as the simpler $B_{(3,1)}(v_0)$
given by Eq. (\ref{eq:108}) that I guessed on my second
attempt without doing any numerical error minimization.
To get a relative accuracy better than roughly
one part in 234 for $b_{\rm crit}(v_0)$, one would need
a different functional form than $B_{(\alpha,\beta)}(v_0)$
given by Eq. (\ref{eq:107}),
but I won't pursue such refinements here.

	Since the critical impact parameter $b_{\rm crit}$
goes as $v_0^{-7/5}$ for small velocity $v_0$ and as
$v_0^{-1/2}$ for large $v_0$,
one can see that if one multiplies $b_{\rm crit}$
by $v_0$ to get the critical initial angular momentum
$L_{\rm crit}(v_0)$
(per unit object mass $M$, and in units of $L_u^2/T_u$
after dividing out by this $M$),
this rises with both low and high $v_0$
and hence must have some minimum in between.
By calculating several points near the minimum,
I found that the minimum value is
 \begin{equation}
 L_{\rm min} \approx 2.9026,
 \label{eq:110}
 \end{equation}
which occurs near the sample point
$(v_0, b_{\rm crit}(v_0)) = (0.834385, 3.48671)$
given above.
When one restores the units and
multiplies by the object mass $M$ to
get the true minimum angular momentum
of the object about the initial static, straight
string configuration, one gets
 \begin{equation}
 L_{\rm min} \approx 2.9026 \left[{\pi^2\over 32}\mu
	\left(1-{Q^2\over M^2}\right)^2\right]^{1/3} M^2
 \approx 1.9611 \mu^{1/3} (1-Q^2/M^2)^{2/3} M^2.
 \label{eq:111}
 \end{equation}
A compact gravitating (and possibly electrically charged)
object approaching from infinity toward
an initially static, straight, infinitely long cosmic string
with angular momentum (about the initial string axis)
less than this would inevitably be captured by the string.

\section{Crude formula for the critical impact parameter
for all velocities}

\hspace{.20in}	In the test-string approximation
\cite{Page},
I calculated the critical impact parameter
for velocities much lower than the speed of light
and then a correction term linear in the velocity
divided by the speed of light, getting Eq. (88)
of that paper.  In the case of a Schwarzschild
black hole ($Q=0$), I arbitrarily added a
(relatively small) constant term
to make the resulting formula agree with the known
critical impact parameter at the speed of light
\cite{JPD}, which is the same as the critical
impact parameter for a massless point particle,
$\sqrt{27}M$.
The resulting formula, Eq. (89) of \cite{Page},
though just a guess by no means rigorously
derived, is apparently a very good approximation
to the actual critical impact parameter
for a test string at all calculated velocities
(i.e., up to a relativistic gamma-factor of $\gamma_0 \sim 10$) \cite{JPD3,JPDprivate},
though it misses an infinite series
of discontinuities that I predicted \cite{Page} occur
at large values of $\gamma_0$
(none seen yet in the analyses of \cite{JPDnew,JPD3}
up to $\gamma_0 \sim 10$).
At velocities at all comparable to the speed of light,
the critical impact parameter apparently
must be calculated numerically
by solving partial differential equations
\cite{JPDnew,JPD3},
which is a much more difficult numerical problem
than using the ordinary differential
equations of \cite{Page} and of the present
paper that are only accurate at velocities
very low compared with the speed of light.

	Here I would like to give an analogue
of Eq. (89) of \cite{Page} that incorporates
(a) Eq. (\ref{eq:109}) as an excellent approximation
to my numerical calculations at all velocities
very low compared with the speed of light,
(b) Eq. (88) of \cite{Page} for the term linear
in velocity that is independent of the string tension,
and also (c) the right limit
for the critical impact parameter at
the speed of light for a Reissner-Nordstrom
black hole (arbitrary $Q^2 \leq M^2$).
This last value is the minimum value
of $\sqrt{-g_{\theta\theta}/g_{00}}$
in the Reissner-Nordstrom metric,
which is
 \begin{equation}
 b_{\rm crit}(v_0=1)
 = \sqrt{(3M+\sqrt{9M^2-8Q^2}\:)^3
	\over 2M+2\sqrt{9M^2-8Q^2}}.
 \label{eq:112}
 \end{equation}

	The resulting crude formula for all velocities
for the critical impact parameter for
the capture of a Reissner-Nordstrom black hole
of mass $M$ and charge $Q$
by an initially straight, static cosmic string
with dimensionless tension $\mu \equiv G\mu \ll 1$ is
that $b_{\rm crit}(v_0)$ is roughly the same as
$b_{\rm guess}(v_0)$, with
 \begin{eqnarray}
 {b_{\rm guess}(v_0) \over M}
	\!\!\!\!&=&\!\!\!\!
 \left({\pi\over 2}\right)^{1\over 2}
	\left(1-{Q^2\over M^2}\right)^{1\over 2}
	v_0^{-{1\over 2}}(1-v_0^{1\over 2})
	\left[1+{\pi^2\over 16}\left({9\over 2}\right)^{2\over 5}
		\left(1-{Q^2\over M^2}\right)^{6\over 5}
		\mu^{12\over 5}v_0^{-{18\over 5}}\right]^{1\over 4}
	\nonumber \\
 &+&\!\!\!\!{\pi\over 4}\left(1-{Q^2\over M^2}
	\right)\mu v_0^{-2}(1-v_0^2)
	\left[1+\left({3125\over 1728}\right)^{3\over 11}
		\left(1-{Q^2\over M^2}\right)^{9\over 11}
		\mu^{18\over 11}v_0^{-{27\over 11}}\right]
			^{-{11\over 15}}
	\nonumber \\
 &-&\!\!\!\!{64\over 15}(1-v_0) + \sqrt{(3+\sqrt{9-8Q^2/M^2})^3
	\over 2+2\sqrt{9-8Q^2/M^2}}.
 \label{eq:113}
 \end{eqnarray}

	The first two terms dominate for $v_0 \ll 1$
(where we have now returned
to units in which the speed of light is $c=1$).
For this highly nonrelativistic velocity
these two terms give essentially the same
as Eq. (\ref{eq:109}), which I found numerically
is accurate to greater than 99.5\% accuracy
for all nonrelativistic velocities when $\mu \ll 1$.
The terms which were added to Eq. (\ref{eq:109})
to get these two terms are essentially constants
(up to correction factors that go to unity
when $\mu$ is taken to zero)
at $v_0 \gg \mu^{2/3}$ and were designed to
make these two terms vanish at $v_0=1$.
The third term on the right hand side of Eq. (\ref{eq:113})
is the contribution of
the term linear in $v_0$ calculated for the test string
in \cite{Page} (though with the recognition that
there might be other such terms missed in that calculation,
and perhaps other terms that are independent of $v_0$
and that go as $v_0^{1/2}$), corrected with a constant
term to make this third term also vanish at $v_0=1$.
The fourth (and last) term on the right hand side
of Eq. (\ref{eq:113}) is independent of $v_0$
and gives the critical impact parameter
at the speed of light, the same as that of a massless
particle impinging upon the Reissner-Nordstrom black hole.

	Although I have no real justification of
Eq. (\ref{eq:113}) for relativistic velocities,
the fact that it was found to be a fairly good
approximation for all velocities calculated so far
(i.e., at least for velocities not too ultrarelativistic)
for Schwarzschild black holes \cite{JPDprivate}
leads me to conjecture that it may be accurate
to within several percent for all velocities 
for Reissner-Nordstrom black holes as well,
assuming only that $\mu \ll 1$.
However, apparently only numerical calculations
could confirm this.

	Following Eq. (90) of \cite{Page},
one might note that this crude guess,
Eq. (\ref{eq:113}),
gives a minimum critical impact parameter of
 \begin{equation}
 {b_{\rm guess \; min} \over M}
	\!\approx\!
 6 \left({\pi\over 15}\right)^{1\over 3}
	\left(1\!-\!{Q^2\over M^2}\right)^{1\over 3}
 - \left({\pi\over 2}\right)^{1\over 2}
	\left(1\!-\!{Q^2\over M^2}\right)^{1\over 2}
  - {64\over 15}  + \sqrt{(3\!+\!\sqrt{9\!-\!8Q^2/M^2})^3
	\over 2\!+\!2\sqrt{9\!-\!8Q^2/M^2}}
 \label{eq:114}
 \end{equation}
at
 \begin{equation}
 v_0
	\approx
 {(225\pi)^{1/3}\over 32}
 \left(1-{Q^2\over M^2}\right)^{1\over 3},
 \label{eq:115}
 \end{equation}
ignoring correction terms of order $\mu$
that for $\mu \ll 1$ would be negligible
in comparison with the $O(1)$ errors
I would expect to be in Eq. (\ref{eq:113}).
It would be interesting to learn from
numerical calculations how the
minimum critical impact parameter
(say divided by $M$ to make it dimensionless),
and the velocity at which it occurs,
varies as a function of $Q^2/M^2$
for a Reissner-Nordstrom black hole.

\section{Effects of a long string on the sun or earth}

\hspace{.20in}	It may be amusing
to conclude this hypothetical
paper with a discussion of what would happen if
a long straight string were impinging upon the sun,
earth, or moon.

	Approximate these objects as spherical,
with mass $M$, charge $Q=0$, and radius $R$,
not to be confused with the magnitude of
${\mathbf R}$ defined in Eq. (\ref{eq:41}).
Then the mathematical analysis above
applies so long as the string
does not get inside the object,
so the separation $r$ between the string
and the center of mass of the object,
the magnitude of ${\mathbf r}$
defined in Eq. (\ref{eq:30}),
must remain greater than $R$.
We shall find that for solar system
objects, this puts us in the stiff-string
regime unless $\mu$ is extremely tiny,
much smaller than $\mu \sim 10^{-6}$
expected for cosmic strings.

	In the critical capture case,
inserting the units $T_u$ and $L_u$
from Eqs. (\ref{eq:32}) and (\ref{eq:33})
into Eqs. (\ref{eq:77b}) and (\ref{eq:77c}),
with $L_0 = b_{\rm crit}v_0$ and with
$b_{\rm crit}(v_0)$ being given by
Eq. (\ref{eq:77}) in this stiff-string regime,
gives the minimum separation $r_m$
and the relative velocity $v_m$
at this separation being given by 
 \begin{equation}
 r_m \approx M \left({9\pi^5\over 2048}\right)^{1/5}
	\left(1-{Q^2\over M^2}\right)^{3/5}
	\mu^{1/5} v_0^{-4/5},
 \label{eq:116}
 \end{equation}
 \begin{equation}
 v_m \approx \left[{8\over 3}\left(1-{Q^2\over M^2}\right)
	\mu^2 v_0^2\right]^{1/5}
 \approx \left[{\pi\mu\over 2}\left(1-{Q^2\over M^2}\right)
	{M\over r_m}\right]^{1/2}.
 \label{eq:117}
 \end{equation}
This expression for $v_m$ is $\sqrt{\pi\mu/4}$
times the speed of a test mass,
with the same charge-to-mass ratio $Q/M$
as the object, that falls in to the surface
of the object from a much lower velocity far away,
because the static force given by Eq. (\ref{eq:14})
between a compact object and a string is $\pi\mu/4$
times that between the object and a copy of itself
at the same separation without the string.

	One can now invert Eq. (\ref{eq:116})
and replace $r_m$ by $R$
to get the initial velocity $v_0$ needed so that the critical
capture occurs with the minimum separation being given by
the radius $R$ of the object:
 \begin{equation}
 v_0 \approx \left({9\pi^5\over 2048}\right)^{1/4}
	\left(1-{Q^2\over M^2}\right)^{3/4}
	\mu^{1/4} \left({M\over R}\right)^{5/4}.
 \label{eq:118}
 \end{equation}
The corresponding critical impact parameter is
 \begin{equation}
 b_{\rm crit} \approx
 M \left({512 \over 9\pi^3}\right)^{1/4}
	\left(1-{Q^2\over M^2}\right)^{-1/4}
	\mu^{1/4} \left({M\over R}\right)^{-7/4}.
 \label{eq:119}
 \end{equation}
For the initial velocity to obey the criterion
$v_0 \ll (1-Q^2/M^2)^{1/3}\mu^{2/3}$ for the stiff-string
approximation to be valid during the critical
capture, one needs
 \begin{equation}
 \left(1-{Q^2\over M^2}\right)\left({M\over R}\right)^3
	\ll \mu \ll 1.
 \label{eq:120}
 \end{equation}
This is an extremely weak restriction for solar system
objects, since $M/R$ is approximately
$2.12\times 10^{-6}$ for the sun,
$6.96\times 10^{-10}$ for the earth, and
$3.14\times 10^{-11}$ for the moon.

	For the sun, with mass
$M \approx 1.47662504$ km,
charge $Q \approx 0$,
and radius $R \approx 6.960\times 10^5$ km,
critical capture with the minimum separation
being $R$ gives the initial velocity and impact
parameter as
$v_0 \approx 0.827(\mu/10^{-6})^{1/4}\; {\rm m/s}
\approx 2.98(\mu/10^{-6})^{1/4}$ km/hr
and $b_{\rm crit}
\approx 4.61\times 10^{11}(\mu/10^{-6})^{1/4}\;{\rm m}
\approx 3.08(\mu/10^{-6})^{1/4}$ AU
(astronomical units).
The relative velocity of the sun and string
when the string just grazes the sun is
$v_m \approx 547(\mu/10^{-6})^{1/2}\; {\rm m/s}
\approx 1970(\mu/10^{-6})^{1/2}$ km/hr.
When the string is grazing the sun,
it gives the sun an acceleration of
$a = \pi\mu M/(4 r^2)
\approx 2.15\times 10^{-4}(\mu/10^{-6})\; {\rm m/s^2}$,
which is less than the average acceleration
of the earth toward the sun,
$5.93\times 10^{-3}\; {\rm m/s^2}$,
unless $\mu \geq 2.76\times 10^{-5}$.
Long slow strings with $\mu \sim 10^{-6}$ or less
apparently don't give much danger of pulling the sun
away from the earth, but they could change
the orbital characteristics of the earth by a few percent,
with possibly disastrous climatic effects.

	For the earth, with $M \approx 0.444$ cm,
$Q \approx 0$, and $R \approx 6371$ km, one gets
$v_0 \approx 3.65\times 10^{-5}
(\mu/10^{-6})^{1/4}\; {\rm m/s}
\approx 13.1(\mu/10^{-6})^{1/4}$ cm/hr,
$b_{\rm crit}
\approx 1.73\times 10^{12} (\mu/10^{-6})^{1/4}\; {\rm m}
\approx 11.6(\mu/10^{-6})^{1/4}$ AU,
and $v_m \approx 9.91(\mu/10^{-6})^{1/2}\; {\rm m/s}
\approx 35.7(\mu/10^{-6})^{1/2}$ km/hr,
if the earth were isolated so that we
could neglect the larger effect of the sun
on the earth during the interaction of 
the string with the earth.
When the string is grazing the earth,
it gives the earth an acceleration of
$a \approx 7.71\times 10^{-6}(\mu/10^{-6})\; {\rm m/s^2}$,
which is less than the average acceleration
of the earth toward the sun if
$\mu < 7.68\times 10^{-4}$.
Thus a string with roughly this small a tension
or smaller could not kidnap the earth from the sun.
The average acceleration of the earth toward
the moon is about $3.32\times 10^{-5}\; {\rm m/s^2}$,
which is greater than the acceleration of the
earth toward a string grazing it if
$\mu < 4.30\times 10^{-6}$,
so a cosmic string with $\mu \sim 10^{-6}$
would be unable to pull the earth away from
either the sun or the moon
(barring contrived resonance effects),
though there is not a large margin of safety
for pulling the earth away from the moon.

	For the moon, with $M \approx 0.0546$ mm,
$Q \approx 0$, and $R \approx 1738$ km, one gets
$v_0 \approx 7.58\times 10^{-7}
(\mu/10^{-6})^{1/4} {\rm m/s}
\approx 2.73(\mu/10^{-6})^{1/4} {\rm mm/hr}$,
$b_{\rm crit}
\approx 5.34\times 10^{12} (\mu/10^{-6})^{1/4} {\rm m} \\
\approx 35.7(\mu/10^{-6})^{1/4}$ AU,
and $v_m \approx 2.10(\mu/10^{-6})^{1/2}\; {\rm m/s}
\approx 7.58(\mu/10^{-6})^{1/2}$ km/hr,
assuming the moon were isolated.
At the moon's surface, a string would give it
an acceleration of
$a \approx 1.27\times 10^{-6}(\mu/10^{-6})\; {\rm m/s^2}$,
which is less than the average acceleration
of the moon toward the sun if
$\mu < 4.66\times 10^{-3}$,
and is less than the average acceleration
of the moon toward the earth,
$2.70\times 10^{-3}\; {\rm m/s^2}$,
if $\mu < 2.12\times 10^{-3}$.
Thus the moon is in even less danger of being
pulled away from the sun than the earth is,
and even if the earth-moon system were isolated
from the sun,
a string with reasonably small tension
could not remove the moon from the earth.

	Another effect one can estimate
is that of a string on the tides.
If the earth had a string grazing it
or running through it, presumably
the curvature of spacetime in and near
the earth would be altered by a fraction
of order $\mu$.
The Riemann curvature tensor at the surface
of the earth has the orthonormal component
$R^0_{\;r0r}
= 2M/R^3 \approx 3.08\times 10^{-6}\; {\rm s}^{-2}
\approx 40.0 \: {\rm hr}^{-2}$,
which is about $1.79\times 10^7$ times the
corresponding curvature contribution from
the moon, and about $3.89\times 10^7$ times the
corresponding curvature contribution from
the sun.
Therefore, if a cosmic string ran through
the earth, the perturbation of the earth's tidal force
would be of the order of $18(\mu/10^{-6})$
times the tidal force of the moon.
If the earth's axis were not aligned with
the direction of the string,
as the string cuts a swath through
the rotating earth, the resulting
perturbation in the earth's tidal (curvature)
gravitational field might produce
Bay-of-Fundy size tides at many coastal cities.
Bids are now being accepted for the rights to
a disaster movie.

\section*{Acknowledgments}

\hspace{.20in}	I thank Jean-Pierre De Villiers
and Valeri Frolov for getting me interested in this subject,
and I thank them and Patrick Brady, Gary Horowitz,
and Brandon Carter for further discussions.
This work was supported in part by the
Natural Sciences and Engineering Research Council
of Canada.


\newpage
\baselineskip 5pt

\begin{thebibliography}{99}

\bibitem{JPD} J.-P. De Villiers and V. Frolov,
Int. J. Mod. Phys. {\bf D7}, 957-67 (1998), gr-qc/9711045.

\bibitem{JPDnew} J.-P. De Villiers and V. Frolov,
Phys. Rev. {\bf D58}, 105018 (1998), gr-qc/9804087.

\bibitem{Page} D. N. Page,
Phys. Rev. {\bf D58}, 105026 (1998), gr-qc/9804088.

\bibitem{JPD3} J.-P. De Villiers and V. Frolov,
``Gravitational Scattering of Cosmic Strings
by Nonrotating Black Holes,''
University of Alberta preprint Thy 08-98 (1998),
gr-qc/9812016.

\bibitem{Nambu} Y. Nambu, in Proceedings
of the International Conference
on {\em Symmetries \& Quark Models},
Wayne State University (1969);
Lectures at the Copenhagen Summer Symposium (1970).

\bibitem{Goto} T. Goto,
Prog. Theor. Phys. {\bf 46}, 1560 (1971).

\bibitem{Carter} B. Carter,
Phys. Lett. {\bf B224}, 61, (1989);
J. Geom. Phys. {\bf 8}, 53 (1992);
Class. Quant. Grav. {\bf 9}, 19 (1992);
Phys. Rev. {\bf D10}, 4835 (1993);
in {\em Formation and Interactions of Topological Defects},
eds. R. Brandenberger and A.-C. Davis
(Plenum, New York, 1995), p. 304, hep-th/9611054;
``Brane Dynamics for Treatment of Cosmic Strings and Vortons,''
Tlaxcala lecture notes, 2nd Mexican School
on Gravitation and Mathematical Physics,
1-7 Dec. 1996,
eds A. Garcia, C. Lammerzahl, A. Macias, and D. Nunez,
hep-th/9705172.

\bibitem{Vil} A. Vilenkin,
Phys. Rev. {\bf D23}, 852 (1981).

\bibitem{Smi} A. G. Smith,
Tufts University Report TUTP-86-11 (1986),
published in G. W. Gibbons, S. W. Hawking,
and T. Vachaspati, eds.,
{\em The Formation and Evolution of Cosmic Strings}
(Cambridge University Press, Cambridge, 1990).

\bibitem{who1} V. P. Frolov, V. D. Skarzhinsky, A. I. Zelnikov,
and O. Heinrich, Phys. Lett. {\bf B224}, 255 (1989);
B. Carter and V. P. Frolov, Class. Quant. Grav. {\bf 6}, 569 (1989).

\bibitem{JPDprivate} J. P. De Villiers,
private communication.

\bibitem{CarterPrivate} B. Carter,
private communication.

\end{thebibliography}
\end{document}